\documentclass{article}

    \PassOptionsToPackage{numbers, compress}{natbib}

\usepackage[final]{neurips_2024}

\usepackage[utf8]{inputenc} 
\usepackage[T1]{fontenc}    

\usepackage{hyperref}       
\usepackage{url}            
\usepackage{booktabs}       
\usepackage{amsfonts}       
\usepackage{nicefrac}       
\usepackage{microtype}      
\usepackage{xcolor}         
\usepackage{array}  
\usepackage{multirow}
\usepackage{tabularx}
\usepackage{graphicx} 
\usepackage{enumitem}
\usepackage{amsmath}
\usepackage{subfigure}
\usepackage{wrapfig}
\usepackage{float}
\usepackage{algorithm}
\usepackage{algorithmic}
\usepackage{microtype}

\newtheorem{definition}{Definition}

\title{Taming the Long Tail in Human Mobility Prediction}

\author{%
Xiaohang~Xu\textsuperscript{1},~~Renhe~Jiang\textsuperscript{1}\thanks{Corresponding author}~~,~~Chuang~Yang\textsuperscript{1},~~Zipei~Fan\textsuperscript{1},~~ 
  Kaoru~Sezaki\textsuperscript{1} \\
  \textsuperscript{1}The University of Tokyo\\
  \texttt{xhxu@g.ecc.u-tokyo.ac.jp}\\
  \texttt{\{jiangrh, chuang.yang\}@csis.u-tokyo.ac.jp}\\
\texttt{\{fanzipei, sezaki\}@iis.u-tokyo.ac.jp}\\
}

\begin{document}

\maketitle

\begin{abstract}
With the popularity of location-based services, human mobility prediction plays a key role in enhancing personalized navigation, optimizing recommendation systems, and facilitating urban mobility and planning.
This involves predicting a user's next POI (point-of-interest) visit using their past visit history. 
However, the uneven distribution of visitations over time and space, namely the long-tail problem in spatial distribution, makes it difficult for AI models to predict those POIs that are less visited by humans.
In light of this issue, we propose the \underline{\bf{Lo}}ng-\underline{\bf{T}}ail Adjusted \underline{\bf{Next}} POI Prediction (LoTNext) framework for mobility prediction, combining a Long-Tailed Graph Adjustment module to reduce the impact of the long-tailed nodes in the user-POI interaction graph and a novel Long-Tailed Loss Adjustment module to adjust loss by logit score and sample weight adjustment strategy. 
Also, we employ the auxiliary prediction task to enhance generalization and accuracy.
Our experiments with two real-world trajectory datasets demonstrate that LoTNext significantly surpasses existing state-of-the-art works.
\end{abstract}

\section{Introduction}
Human mobility prediction is essential in various applications, aiming to forecast the next Point of Interest (POI) a user may visit based on their historical location data, preferences, and patterns~\cite{huang2019variational,han2021point,zhu2023difftraj,wang2024large}. 
By predicting user movements, it supports urban planning, traffic management, and environmental protection, and provides intelligent personalized Location-Based Social Networking (LBSN) services~\cite{gao2022self,wu2020personalized}, enhancing people's life quality.

The growth of POI prediction tasks is closely linked to the development of LBSN platforms, where users frequently share their itineraries and reviews, leading to a substantial accumulation of geographical visitation data.
However, data collection faces challenges due to network and privacy constraints on mobile devices and the requirement for user authorization to record check-ins. 
This often results in data being sparse and biased towards popular locations, exhibiting a severe long-tail effect.
Currently, these methods fall into two primary categories: \textit{Sequence-based} and \textit{Graph-based} models.
\begin{itemize}[leftmargin=*]
\item \textit{Sequence-based} models treat users' trajectories as independent visitation sequences. Existing methods include Recurrent Neural Networks (RNNs)~\cite{feng2018deepmove,jiang2018deepurbanmomentum,jiang2018deep}, Long Short Term Memory (LSTM)~\cite{chen2020dualsin,liu2022real,jiang2022will,jiang2021transfer} and Gated Recurrent Unit (GRU)~\cite{fan2018online,fan2022online} for modeling the rich spatial-temporal information implied in the visitation sequence.
\item \textit{Graph-based} models focus on building models and data structures to capture trend information in the data to enhance the prediction performance, such as the movement trends among all users
~\cite{yang2022getnext,xu2023revisiting,wang2023eedn,wang2023adaptive,yan2023spatio}, geographic adjacency~\cite{rao2022graph,luo2023timestamps,qin2023diffusion}, and category transition between POIs~\cite{zhang2020modeling,yu2020category}. This helps in modeling complex global visitation preferences and the semantic context of locations.
\end{itemize}

\begin{wrapfigure}{r}{0.5\textwidth}
    \centering
    \vspace{-0.2cm}
    \includegraphics[width=0.5\textwidth]{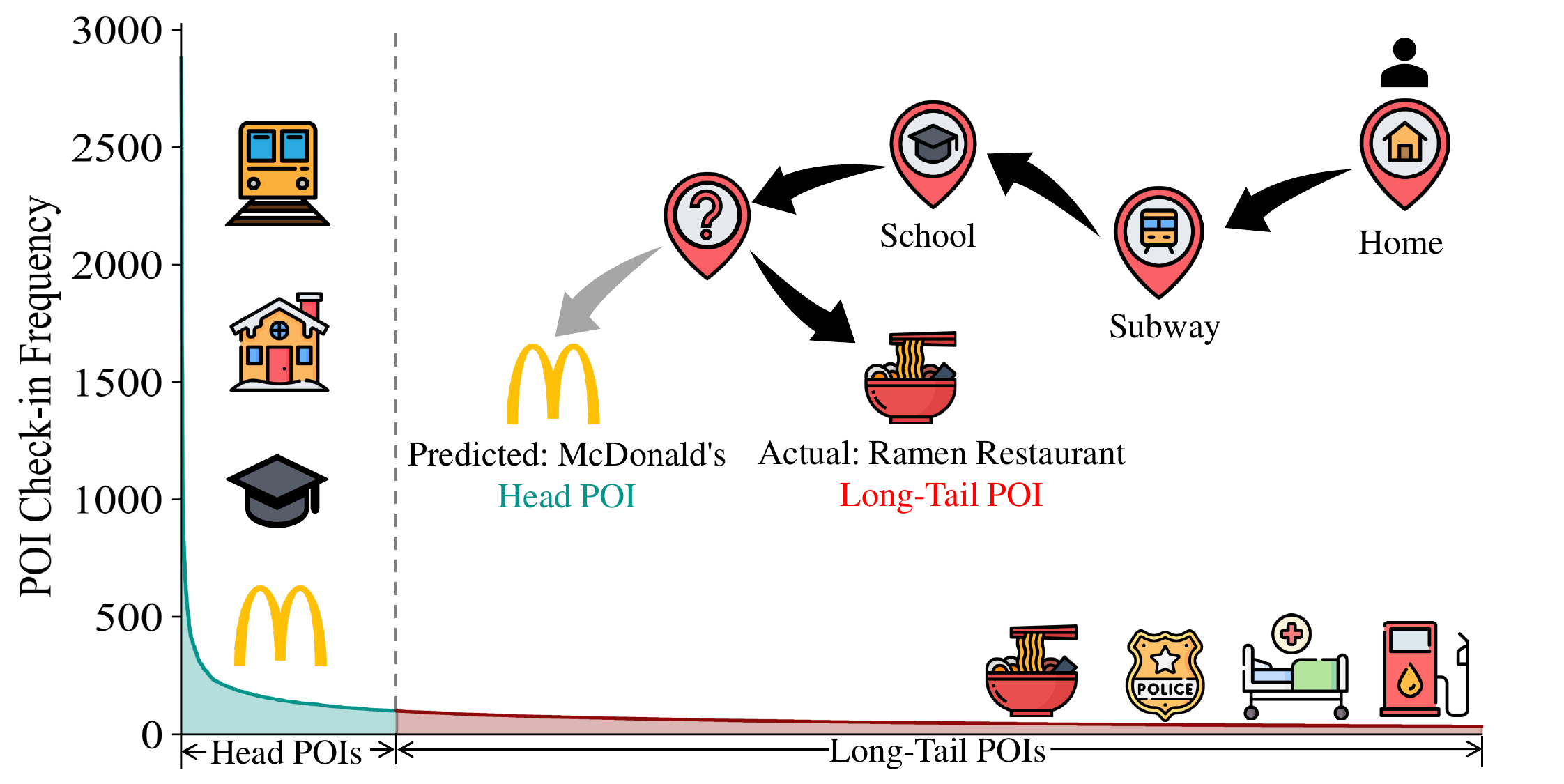}
    \vspace{-0.3cm}
    \caption{The long-tailed distribution for POI check-in frequency from the Gowalla dataset.}
    \label{fig:longtail}
    \vspace{-0.3cm}
\end{wrapfigure}

Nevertheless, existing works often overlook the intrinsic long-tailed distribution problem in spatial visitation patterns. 
As shown in Figure~\ref{fig:longtail}, it provides the evidence of long-tailed distribution on Gowalla\footnote[1]{\url{https://snap.stanford.edu/data/loc-gowalla.html}} dataset, a real LBSN dataset.    
From the visualization results, it is evident that only a few POIs are visited more than 100 times.
In addition, the illustrative diagram above the graph presents a hypothetical scenario. 
The prediction model might inaccurately predict that a user would visit a common location, such as McDonald's (a Head POI), while the user actually visits a less common place like a ramen restaurant (a Long-Tail POI). This highlights the importance of designing models capable of accurately predicting visits to Long-Tail POIs.
The concept of a long-tailed distribution, while first extensively studied and addressed within the computer vision (CV) field~\cite{yang2022survey}, manifests differently in the context of POI prediction. 
In POI prediction, long-tailed POIs are embedded within users' complex trajectories. 
This distinction means that unlike in CV, where long-tailed samples can be selectively augmented to balance datasets, selecting long-tailed POIs without considering the spatial-temporal context in which they occur risks losing crucial trajectory information.

Against this background, to mitigate the long-tail problem in the human next POI prediction task, we propose the \underline{\bf{Lo}}ng-\underline{\bf{T}}ail Adjusted \underline{\bf{Next}} POI prediction (LoTNext) \footnote[2]{\url{https://github.com/Yukayo/LoTNext}} framework, which is a generic framework aimed at optimizing and fully utilizing long-tailed POI information. 
More specifically, our solution first employs a Long-Tailed Graph Adjustment module to reduce noise and long-tailed nodes in the user-POI interaction graph, thereby mitigating the impact of long-tailed POIs on model performance. 
Through graph adjustment, the model can more accurately capture spatial-temporal information from trajectory contexts. 
Furthermore, to prevent the model from overly focusing on head POIs (high-frequency POIs), we propose the Long-Tailed Loss Adjustment module to balance the loss between head and tail POI data. 
Finally, to alleviate the intrinsic sparsity issue without introducing additional data sources, we incorporate auxiliary prediction tasks to further integrate the POI feature and spatial-temporal information. 

We conclude our contributions as follows: (1) We propose the LoTNext framework based on graph adjustment to effectively address the challenges of dataset-inherent sparsity in the user-POI interaction graph.
(2) We design the Long-Tailed Loss Adjustment module for adaptive sample re-weighting, which more effectively balances the loss between head and tail samples. 
(3) We introduce the auxiliary prediction task, which achieves complementarity of POI feature information and spatial-temporal information.
(4) We evaluate LoTNext on two public LBSN datasets, comparing it with numerous baselines. 
The results demonstrate that LoTNext significantly outperforms state-of-the-art methods.

\section{Related Work}
\textbf{Next POI Prediction.} Most current works on the next POI prediction treat trajectories as time series, further incorporating spatial-temporal contexts into models to enrich the semantics of POIs. 
The pioneering ST-RNN~\cite{liu2016predicting} introduces spatial-temporal intervals to RNN for context awareness. 
DeepMove~\cite{feng2018deepmove} integrates LSTM with attention mechanisms to consider both the short-term and long-term preferences of users comprehensively, and LSTPM~\cite{sun2020go} enhances spatial context integration. 
The Flashback model~\cite{yang2020location} tackles user sparsity by mining similar contexts in historical data. 
However, due to the limited capability of RNN in modeling long sequences, researchers have explored using graphs for improvements. 
GETNext~\cite{yang2022getnext}, based on the Transformer architecture, combines global mobility patterns graph with various spatial-temporal contexts to fully utilize information among similar user trajectories for improving prediction performance. 
Graph-Flashback~\cite{rao2022graph} considers constructing a knowledge graph to improve POI representation and integrates it with sequence recommendation models. 
SNPM~\cite{yin2023next} builds a POI similarity graph to aggregate similar POIs and enhance POI representation results. 
However, all these studies overlook the significant impact of the long-tail problem on the next POI prediction.

\textbf{Long-Tailed Learning.} The long-tail problem has always been a focus in the fields of CV~\cite{du2023no,tao2023local} and recommendation systems~\cite{kim2019sequential}. 
The most direct solution is re-sampling~\cite{zhang2023deep}, varying from random to progressively balanced.
Another common strategy is logit adjustment~\cite{MenonJRJVK21,provost2000machine,tian2020posterior}, aimed at modifying the logistic output to address the imbalanced data class problem.
A study by Google~\cite{MenonJRJVK21} has proven that logit adjustment satisfies Fisher consistency and can effectively minimize the average error per category.
Compared to the CV field, the long-tail problem is more pronounced in recommendation systems~\cite{beutel2017beyond,liu2020long}. 
Typically, the number of items far exceeds the number of users, leading to many items rarely or infrequently accessed by users. 
Some works tackle the long-tail problem by learning item similarities through random walk algorithms~\cite{YinCLYC12} or utilizing transfer learning to transfer the knowledge from head to tail data~\cite{zhang2021model}. 
\cite{wei2023meta} used meta-learning to enhance the information representation of the user-item graph. 
\cite{luo2023improving} introduced a novel edge addition module to enrich the connectivity for tail samples. 
However, unlike traditional recommendation tasks, the next POI prediction involves complex spatial-temporal semantics due to the nature of trajectory data, making it more challenging to improve the representation of long-tailed POI samples. 
\emph{To the best of our knowledge, our work is the first to propose a general framework for the next POI prediction under the long-tail problem.}

\section{Problem Definition}
Given a user set $U$ = $\{u_1, u_2, ..., u_{|U|}\}$ and a POI set $P$ = $\{p_1, p_2, ..., p_{|P|}\}$, with $|U|$ and $|P|$ indicating the number of users and POIs respectively, 
we denote the POI check-in as a triplet $\langle u, p, t \rangle$, which means a user $u$ visits POI $p$ at time $t$. Each POI $p$ is a triplet $p$ = $\langle lat, lon, freq \rangle$, representing its latitude, longitude, and visit frequency. We proceed to outline our problem definition as follows.
\begin{definition}[User Next POI Prediction] 
Given a user check-in sequence denoted as $Q_{u}$= $(\langle p_1, t_1 \rangle, \langle p_2, t_2 \rangle, \dots, \langle p_n, t_n \rangle)$, our goal is to predict a list of top POIs that the user $u$ is likely to visit next, which can be taken as a typical sequence classification task over $|P|$ POI candidates. In particular, our work focuses on how to accurately predict the ``less visited'' POIs belonging to the long-tailed interval. 
\end{definition}

\begin{figure*}[!t]
\centering	
\includegraphics[width=\textwidth]{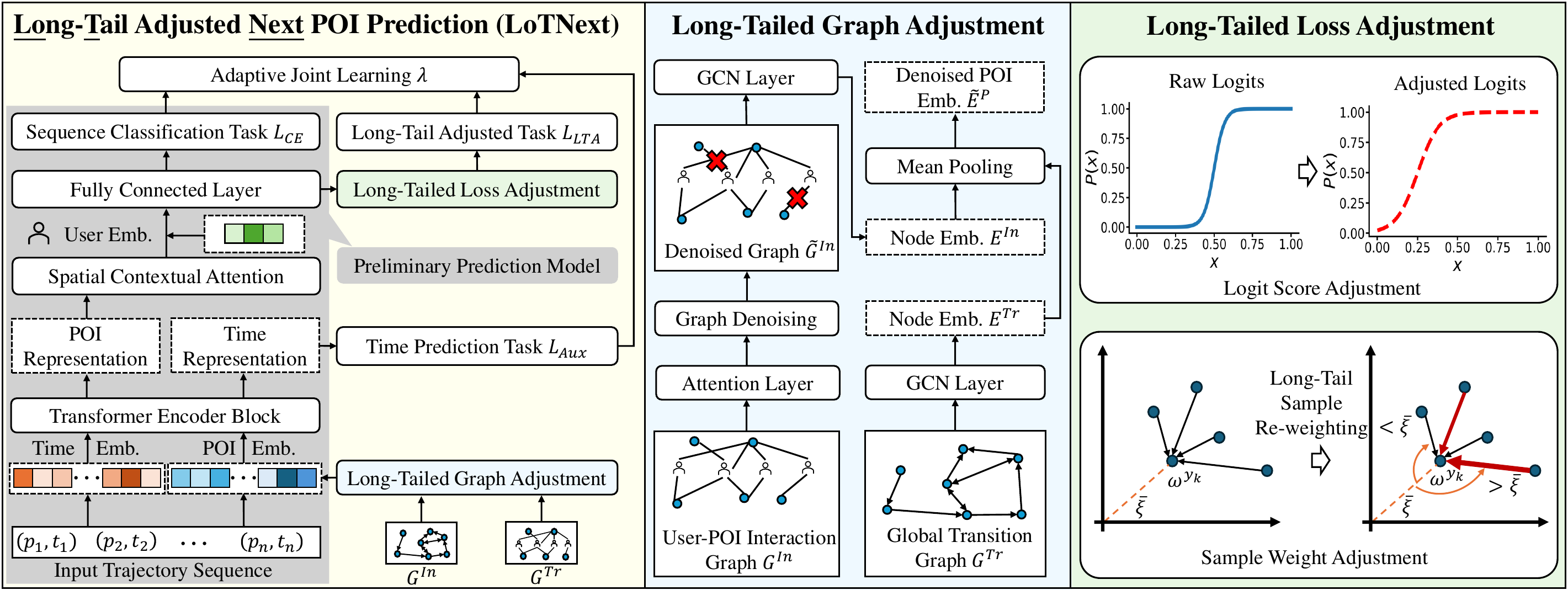}
\caption{The Architecture of \underline{\textbf{Lo}}ng-\underline{\textbf{T}}ail Adjusted Network for \underline{\textbf{Next}} POI Prediction \textbf{(LoTNext)}.}
\label{fig:gdla}
\end{figure*}

\section{Methodology}
In this section, we introduce the details of the LoTNext framework, as shown in Figure~\ref{fig:gdla}, which consists of the preliminary POI prediction model, Long-Tailed Graph Adjustment module, and Long-Tailed Loss Adjustment module. 
\subsection{Preliminary Model}
We first construct a preliminary end-to-end model, which is designed for precise next POI prediction.

\noindent\textbf{Trajectory Embedding Layer.} 
For the embedding generation, we initialize embeddings for POIs $E^{P} \in \mathbb{R}^{|P| \times d_p}$, timestamps $E^{T} \in \mathbb{R}^{|T| \times d_t}$, and users $E^{U} \in \mathbb{R}^{|U| \times d_u}$, where $d_p$, $d_t$ and $d_u$ are the corresponding embedding dimension and $|T|$ is the number of the time slots.
In our case, considering each hour slot over a week, there are 168 time slots in total. 
During the sequence processing phase, we select embedding from \({E}^{P}\) and \({E}^{T}\) based on the POI and time indices in the input sequence $Q_u$ to construct an embedding sequence $X \in \mathbb{R}^{n \times (d_p+d_t)}$, as $X=\left[({E}^{P}_{p_1}, {E}^{P}_{p_2}, ..., {E}^{P}_{p_n}) || (E^{T}_{t_1}, E^{T}_{t_2}, ..., E^{T}_{t_n})\right]$, 
where $n$ is the sequence length and $||$ denotes the concatenation operation.

\noindent\textbf{Transformer Encoder.} 
Transformer~\cite{vaswani2017attention} architecture in modeling trajectories has been demonstrated in multiple studies~\cite{xu2023revisiting,yang2022getnext,xue2021mobtcast}, we adopt its encoder block to encode spatial-temporal contexts within trajectories, focusing on capturing long-distance dependencies through multi-layer stacking.
To maintain positional information in the sequence, we incorporate a learnable positional embedding $E_{pos} \in \mathbb{R}^{n \times d_{p}}$ with the raw embedding sequence $X$ to form the Transformer input $\widetilde{X}= X + E_{pos} $.
Next, we define the Transformer encoder block as follows:
\begin{equation}
\begin{aligned}
Z &= \text{LayerNorm}(\widetilde{X} + \text{Multi-Head Attention}(\widetilde{X})), \\
\widetilde{Z} &= \text{LayerNorm}(Z) + \text{FFN}(Z)), 
\end{aligned}
\label{eq:3}
\end{equation}
where FFN is a fully connected layer and the Multi-Head Attention can be described as:
\begin{equation}
\begin{aligned}
\text{Multi-Head}(\widetilde{X})&=[\text{head}_{1}||\text{head}_{2}||...||\text{head}_{h}]W^{O}, \\
\text{head}_{i}&=\text{Softmax} \left( \frac{\widetilde{X}^iW^{Q}(\widetilde{X}^iW^{K})^{T}}{\sqrt{d_{k}} } \right)\widetilde{X}^iW^{V},
\end{aligned}
\label{eq:4}
\end{equation}
where $\widetilde{X}^{i}$ is the input of the $i$-th head, $W^{O},W^{Q},W^{K}$, and $W^{V}$ are the learnable weights matrix, $h$ is the number of the head, and $\sqrt{d_k}$ is the scaling factor.

\noindent\textbf{Spatial Contextual Attention Layer.}
Inspired by Flashback~\cite{yang2020location}, we introduce the Spatial Contextual Attention Layer to analyze the relationship between spatial proximity and user interactions. 
It assigns dynamic weights to POIs in a sequence, taking into account both the order and geographical distances, focusing on POIs most influential to future movements.
The spatial weight $\omega_{k}$ for each POI $p_k$ in the sequence $(p_1, p_2, ..., p_k, ..., p_n)$, considering its spatial distance to all previous POIs $p_1$$\sim$$p_k$, is:
\begin{equation}
\omega_{k} = \sum_{j=1}^{k} \left( e^{-\beta(\Delta(p_j, p_k))} + \epsilon \right),
\label{eq:5}
\end{equation}
where $\Delta(p_j, p_k)$ is the haversine distance between $p_j$ and $p_k$, $\beta$ is the distance decay weight, and $\epsilon$ is a small constant to prevent division by zero.
Considering $\tilde{z}_k$, an element from the Transformer output sequence $\widetilde{Z} = (\tilde{z}_1, \tilde{z}_2, ..., \tilde{z}_k, ..., \tilde{z}_n)$. 
The refined output $\tilde{z}'_{k}$ is obtained by applying spatial weight to the cumulative previous outputs, defined as follows:
\begin{equation}
\tilde{z}'_{k} = \frac{\sum_{j=1}^{k} \omega_{j} \cdot \tilde{{z}}_{j}}{\sum_{j=1}^{k} \omega_{j}}.
\label{eq:6}
\end{equation}
\noindent\textbf{Prediction Layer.}
To provide personalized predictive outcomes and ensure accurate representation even for users with fewer check-ins, we further introduce user embeddings $E^{U}_{u}$ and fuse it with refined output $\widetilde{Z}'$ to form the input $\mathcal{O}$ = $[\widetilde{Z}'||E^{U}_{u}]$ for the final fully connected layer $L=\mathcal{O}W+b^P$,
where $W \in \mathbb{R}^{(d_p+d_u) \times |P|}$ is the weight matrix of the fully connected layer, $b \in \mathbb{R}^{|P|}$ is the bias, and $L \in \mathbb{R}^{n \times |P|}$ is the logit scores for $n$ steps of POI prediction.  
As the POI prediction is essentially a sequence classification task, we adopt the standard cross-entropy loss $\mathcal{L}_{CE}$ as follows:
\begin{equation}
\mathcal{L}_{CE} = -\frac{1}{N}\sum_{k=1}^{N} \sum_{i=1}^{|P|}y_{i}^{k} \log \left( \frac{\exp(l_{i}^{k})}{\sum_{j=1}^{|P|} \exp(l_{j}^{k})} \right),
\label{eq:18_2}
\end{equation}
where $l^k_i$ $\in$ $\mathbb{R}^1$ represents the logit score of the \(k\)-th sample for the \(i\)-th POI candidate in $P$, and $y_{i}^{k}$ is the ground-truth indicator on POI label $i$ for the $k$-th sample.
In our implementation, we mix the $n$ steps of prediction and $B$ samples in one batch together as \(N=n \times B\) samples in total.   

\subsection{Long-Tailed Graph Adjustment}
In the next POI prediction task, we model user-POI interactions via a User-POI Interaction Graph $\mathcal{G}^{In}$ = $(\mathcal{V}^{In}, \mathcal{A}^{In})$, where $\mathcal{V}^{In}$ = $[E^U || E^P]$ $\in$ $\mathbb{R}^{(|U|+|P|) \times d}$ is the input node feature matrix of the $\mathcal{G}^{In}$, $d$ = $d_p$ = $d_u$, and $ \mathcal{A}^{In}$ $\in$ $\mathbb{R}^{|U| \times |P|}$ is the adjacent matrix. 
It's a bipartite graph where user $U$ and POI $P$ nodes connect through edges symbolizing interaction frequencies or preferences. 
Graph Neural Networks (GNNs)~\cite{wu2020comprehensive} can leverage graphs to learn complex node representations, but performance hinges on graph quality. 
However, the $\mathcal{G}^{In}$ often has long-tailed distributions—most interactions are limited to few nodes with high visit frequency, which affects the quality of node embeddings and model efficacy.
To tackle the long-tail problem in $\mathcal{G}^{In}$, we propose a denoising layer to prune and reduce sparse interactions caused by the distribution. 
This layer evaluates edge importance, retaining only beneficial edges for learning. 
Initially, an attention layer weights edges according to user-POI embedding interactions, processed by a multilayer perceptron (MLP) to obtain attention scores:
\begin{equation}
A_{ij} = \sigma(W^B \cdot \text{LeakyReLU}(W^A [E^{U}_{i}|| E^{P}_{j}] + b^A) + b^B). \\
\label{eq:1}
\end{equation}
Here, $\sigma$ denotes the sigmoid function, ensuring that the attention scores $A_{ij}$ lie in the (0, 1) interval, $E^{U}_{i}$ and $E^{P}_{j}$ means the embedding of user and POI, $W$ represents the trainable weight matrix, and $b$ represents the bias.
Based on the attention score $A_{ij}$, the denoising process applies a thresholding operation to filter out edges with scores below a threshold $\delta$, effectively reducing noise and focusing on high-quality interactions. 
This process aims to derive the denoised graph $\mathcal{\widetilde{G}}^{In}=(\mathcal{V}^{In}, \mathcal{\widetilde{A}}^{In})$ can be formalized as: 
\begin{equation}
\mathcal{\widetilde{A}}_{ij}^{In} = \mathcal{A}_{ij}^{In} \cdot \mathbf{1}[A_{ij} \geq \delta],
\label{eq:2_1}
\end{equation}
where $\mathcal{\widetilde{A}}_{ij}^{In}$ denotes the refined edge and $\mathbf{1}[\cdot]$ is the indicator function. 
The threshold \( \delta \) controls the sparsity of the graph, only edges with weights signifying a strong user-POI relationship are retained in \( \mathcal{\widetilde{G}}^{In} \). 
It is worth noting that when all edges fall below $\delta$, the edge with the highest attention score is retained to prevent isolated nodes in the graph.
The model then leverages the Graph Convolutional Network (GCN)~\cite{DBLP:conf/iclr/KipfW17} layer to learn the node embedding ${E}^{In}$ of the $\mathcal{\widetilde{G}}^{In}$, as follows:
\begin{equation}
{E}^{In} = \text{LeakyReLU}\left((D^{In})^{-\frac{1}{2}}\widetilde{\mathcal{A}}^{In}(D^{In})^{-\frac{1}{2}}\mathcal{V}^{In}W^{In}\right),
\label{eq:2_2}
\end{equation}
where $D^{In}$ is the degree matrix of the $\widetilde{\mathcal{A}}^{In}$, and $W^{In}$ is the graph convolution weight.
It is noted that here we perform a slicing operation $E^{In}$ = $E^{In}[|P|:]$ to select the node embedding representing the POI of $\mathcal{\widetilde{G}}^{In}$.
Beyond merely focusing on direct interactions between users and POIs, we further extend our exploration to utilize all users' check-in data to uncover global mobility patterns among POIs.
We build a user-independent directed Global Transition Graph $\mathcal{G}^{Tr}$ = $(\mathcal{V}^{Tr}, \mathcal{A}^{Tr})$, where $\mathcal{V}^{Tr}$ $\in$ $\mathbb{R}^{|P| \times d_p}$ and $ \mathcal{A}^{Tr} \in \mathbb{R}^{|P| \times |P|}$.
Here, $\mathcal{V}^{Tr}$ is equal to $E^P$, and $\mathcal{A}^{Tr}$ stores the visit frequency between two different POIs.
It is important to note that we do not perform a denoising process on the $\mathcal{G}^{Tr}$, as it accurately reflects the mobility patterns of all users, containing a wealth of global transition information. 
Similarly, we employ GCN refer to Equation~(\ref{eq:2_2}) to learn the node embedding ${E}^{Tr}$ of the $\mathcal{G}^{Tr}$.
Finally, we perform mean pooling to combine the two node embeddings ${E}^{In}$ and ${E}^{Tr}$, which yields the denoised POI embedding $\widetilde{E}^{P}$= $\frac{1}{2}({E}^{In}+{E}^{Tr})$ that incorporate comprehensive user mobility patterns from interaction and transition graphs.
To introduce denoised embedding in our model, we refine our input embedding sequence $X$ construction process as $X=\left[(\widetilde{E}^{P}_{p_1}, \widetilde{E}^{P}_{p_2}, ..., \widetilde{E}^{P}_{p_n}) || (E^{T}_{t_1}, E^{T}_{t_2}, ..., E^{T}_{t_n})\right]$.
\subsection{Long-Tailed Loss Adjustment}
\noindent\textbf{Logit Score Adjustment.}
Traditional classification models often mechanically employ the softmax function for outputting predictions, which may lead to an oversight of the potential discrepancies in the posterior distributions between training and testing data. 
To improve model discrimination, logit adjustment has been explored, which originates in the domain of face recognition~\cite{ren2020balanced,zhao2022adaptive}, 
It involves modifying the model's output layer (i.e., logits) to encourage the generation of more compact intra-class representations while increasing the distance between classes, thereby augmenting the model's capability to handle long-tailed data.

To address the long-tail problem in human next POI prediction tasks, we propose the Logit Score Adjustment module. 
It adjusts the logits by a factor that is inversely correlated with the frequency of occurrence of each label, effectively dampening the influence of frequently occurring labels and amplifying that of rarer ones. 
The adjustment factor $\alpha_i$ for label $i$ with frequency $freq$ is given by:
\begin{equation}
\alpha_i = \tau \left[1 - \frac{\log(freq_i + \epsilon)}{\log(freq_{max} + \epsilon)}\right],
\label{eq:10}
\end{equation}
where $freq_{max}$ is the maximum label frequency observed in the dataset, $\tau$ is the logit adjustment weight and $\epsilon$ is a small constant to stabilize the logarithm operation. 
We can adjust final logits $\widetilde{l}_i \in \mathbb{R}^{1}$ based on the logits $l_i \in \mathbb{R}^{1}$ as $\widetilde{l}_i = l_i + \alpha_i$.

\noindent\textbf{Sample Weight Adjustment.}
Based on the Equation~(\ref{eq:18_2}), for the standard cross-entropy loss, we can find due to the nature of the softmax function, which normalizes the logits $l_{i}^k$ into probabilities, the model can become biased toward head classes. 
This imbalance means that the model's updates are predominantly driven by the head classes, as the loss from incorrectly classified examples in long-tailed classes contributes insignificantly to the overall loss. 
Even marginal improvements in the predictions for these long-tailed classes may contribute insignificantly to the overall loss.
Therefore, it is necessary to reweight long-tailed samples, like with Focal Loss~\cite{lin2017focal}, which reduces the weights of well-classified samples to better focus on minority classes, but it does not explicitly consider the imbalance degree between classes in the long-tailed distribution. 
Unlike Focal Loss, we propose a novel Long-Tail Adjusted (LTA) loss to adaptively re-weight long-tailed samples. Specifically, for the final prediction layer, 
we have the hidden inputs of $N$ samples $\mathcal{O}$ = $(o^1, o^2, ..., o^N)$ and the weights $W$ = $(w^1, w^2, ..., w^{|P|})$ for $|P|$ candidates, where $o^k$ $\in$ $\mathbb{R}^{(d_u+d_p)}$ is from the $k$-th sample. The true class label for the $k$-th sample is denoted by $y_k$. We can take $w^{y_k} \in \mathbb{R}^{(d_u+d_p)}$ as the class ``center'' for the class to which the $k$-th sample truly belongs. Then we assess the impact posed by the $k$-th sample to the overall prediction through the cosine similarity between $o^k$ and $w^{y_k}$ as follows:
\begin{equation}
cos(o^{k},{w}^{y_k}) = \frac{o^k \cdot w^{y_k}}{\|o^k\| \|w^{y_k}\|}.
\label{eq:12}
\end{equation}
Based on these cosine similarities, we compute the adjusted vector magnitude $\xi^{k}$ for each sample as: 
\begin{equation}
\xi^{k}=\left\{\begin{matrix}
  1,\qquad\qquad\qquad\;\,\, & cos(o^k, w^{y_k})> 0,\\
  1-cos(o^k, w^{y_k}), & \; cos(o^k, w^{y_k})\le0.
\end{matrix}\right.
\label{eq:12_1}
\end{equation}
Then we determine the geometric mean of the vector magnitude to serve as a baseline magnitude $\bar{\xi}$.
The traditional definition of the geometric mean of the vector magnitudes is the $N$-th root of their product $\bar{\xi} =  \sqrt[N]{\xi^1\xi^2\cdots\xi^N}$. 
However, it can be problematic in practice due to numerical underflow or overflow when dealing with very small or very large values. 
To mitigate this issue, we utilize logarithm to turn the product into a sum, making the calculation more numerically stable, as follows:
\begin{equation}
\bar{\xi} = \exp\left(\frac{1}{N} \sum_{k=1}^{N} \log(\xi^{k} + \epsilon)\right).
\label{eq:13}
\end{equation}
We calculate adaptive weights $\phi^k$ for each sample using the deviation of vector magnitude from the geometric mean:
\begin{equation}
\phi^k=\left\{\begin{matrix}
  1,\qquad\quad\;\;\;\,\, & \xi^{k}-\bar{\xi}\le 0,\\
  1+\xi^{k}-\bar{\xi}, & \; \xi^{k}-\bar{\xi}>0.
\end{matrix}\right.
\label{eq:14}
\end{equation}
:Finally, the overall Long-Tail Adjusted loss $\mathcal{L}_{LTA}$ can be formulated as:
\begin{equation}
\mathcal{L}_{LTA} = - \frac{1}{N}\sum_{k=1}^{N} \phi^k \sum_{i=1}^{|P|} y_{i}^k \log \left( \frac{\exp(\widetilde{l}_{i}^{k})}{\sum_{j=1}^{|P|} \exp(\widetilde{l}_{j}^k)} \right).
\label{eq:15_2}
\end{equation}
By combining the Logit Score Adjustment and the Sample Weight Adjustment, we present a nuanced approach to recalibrating the model's focus across the spectrum of label frequencies. 
It ensures that each sample contributes to the model's learning process in proportion to its significance, as dictated by the distributional characteristics of the dataset and the discriminative capacity of the model.

\subsection{Model Optimization}
Building upon our Long-Tailed Loss Adjustment module, we further embrace auxiliary prediction tasks to optimize LoTNext. 
To incorporate these tasks, we define a joint loss function that combines three distinct loss components: the standard cross-entropy loss ($\mathcal{L}_{CE}$), the Long-Tail Adjusted Loss ($\mathcal{L}_{LTA}$), and the Mean Squared Error loss for auxiliary time prediction ($\mathcal{L}_{Aux}$). 
Each component serves a critical role: $\mathcal{L}_{CE}$ ensures the fidelity of the next POI prediction, $\mathcal{L}_{LTA}$ addresses the long-tailed data imbalance through adaptive weighting, and $\mathcal{L}_{Aux}$ measures the accuracy of the timing predictions, an auxiliary task that supports the model by providing it with temporal context, thereby improving prediction accuracy and robustness, which can be denoted as:
\begin{equation}
\mathcal{L}_{Aux} = \frac{1}{N}\sum_{k=1}^{N}||\hat{t}^{k} - t^{k}||^{2},
\label{eq:17}
\end{equation}
where $\hat{t}^{k}$ is the forecasted time slot of $k$-th candidate POI and $t^{k}$ is the ground truth time slot.
The overall loss function is constructed as a weighted sum of these components, with the weights $\lambda$ being learnable parameters, as follows:
\begin{equation}
\mathcal{L}_{Joint} = \lambda_{1} \mathcal{L}_{CE} + \lambda_{2} \mathcal{L}_{LTA} + \lambda_{3} \mathcal{L}_{Aux}.
\label{eq:19}
\end{equation}

\section{Experiments}
\label{sec:experiment} 
\noindent\textbf{Datasets \& Baselines.}
We evaluate our LoTNext on two publicly available real-world LBSN datasets: Gowalla and Foursquare\footnote[2]{\url{https://sites.google.com/site/yangdingqi/home/foursquare-dataset}}
Each user check-in record includes the User ID, POI ID, latitude, longitude, and timestamp. To focus solely on the impact of long-tailed POIs and ensure the dataset's quality, we filter out inactive users with fewer than 100 check-ins. 
We then split each user's check-in records according to temporal order, using the first 80\% for training and the remaining 20\% for testing. 
To batch training, we uniformly segment the length of each input trajectory (e.g., 20). 
The specific statistical results are shown in Table~\ref{tab:data-statstic}, where we additionally calculated the percentage of POIs with a frequency smaller than 200 times and smaller than 100 times out of the total number of POIs. 
For instance, defining long-tailed POIs as those with a frequency of less than 100 times, approximately 98.38\% of POIs could be considered long-tailed POIs.
Considering both Table~\ref{tab:data-statstic} and Table~\ref{tab:result}, the reason why the model performs about 20\% points better on Foursquare compared to Gowalla is due to the more severe long-tail effect on the Gowalla dataset, along with a sparser density of the dataset.

\begin{table}[!ht]

\centering
\caption{Basic dataset statistics.}
\resizebox{0.6\textwidth}{!}{
\begin{tabular}{ccc}
\toprule
\textbf{Dataset}               & \textbf{Gowalla} & \textbf{Foursquare} \\ \midrule
Duration                       & 2009.02-2010.10  & 2012.04-2014.01     \\
\#Users                        & 7,768            & 45,343              \\
\#POIs                         & 106,994          & 68,879              \\
\#Check-ins                    & 1,823,598        & 9,361,228           \\
\#Trajectories                 & 84,357           & 429,071             \\ \midrule
Density                        & 0.002194                 & 0.002997                    \\
POI frequency \textless 200 (\%) & 99.57\%           & 89.26\%             \\
POI frequency \textless 100 (\%) & 98.38\%           & 63.70\%             \\ \bottomrule
\end{tabular}
}
\label{tab:data-statstic}
\end{table}
To demonstrate the performance of the LoTNext, we implement the following 10 state-of-the-art methods as the comparison baselines: 
\begin{itemize}[leftmargin=*]
\item \textbf{ST-RNN}~\cite{liu2016predicting} extends the RNN by introducing the spatial and temporal transition matrices.

\item \textbf{DeepMove}~\cite{feng2018deepmove} considers long-term and short-term interests of users by attention mechanism.

\item \textbf{LBSN2Vec}~\cite{yang2019revisiting} introduces the hypergraph and calculates the similarity of users and time embeddings to rank POIs.

\item \textbf{LightGCN}~\cite{he2020lightgcn} simplifies the structure of Graph Convolutional Network (GCN) to learn user preferences for POIs.

\item \textbf{LSTPM}~\cite{sun2020go} proposes geo-nonlocal LSTM to further extend DeepMove structure.

\item \textbf{Flashback}~\cite{yang2020location} searches the most similar hidden states in historical information based on the current context information and updates the model.

\item \textbf{STAN}~\cite{luo2021stan} explores the influence between non-adjacent check-in records in trajectory sequences through the attention mechanism.

\item \textbf{GETNext}~\cite{yang2022getnext} introduces the global mobility patterns of all users into the Transformer architecture to improve model prediction effects.

\item \textbf{Graph-Flashback}~\cite{rao2022graph} combines Spatial-Temporal Knowledge Graph with the sequential model to enrich the representation of each POI.

\item \textbf{SNPM}~\cite{yin2023next} learns the general characteristics of POIs by constructing a POI similarity graph and aggregating similar POIs.
\end{itemize}

\noindent\textbf{Metrics.} To evaluate the model performance, we utilize two of the most common metrics for the next POI prediction: Accuracy@k (Acc@k) and Mean Reciprocal Rank (MRR). 
Acc@k effectively measures whether the true label is present within the top-k predicted results. 
Here, we consider k=1, 5, and 10 to comprehensively assess the model's performance. 
MRR directly quantifies the average rank of the correct label among all predictions when the correct label is not within the top-k predictions, with higher values indicating better average prediction performance by the model.

\noindent\textbf{Settings.} We implement LoTNext using PyTorch 1.13.1 on a Linux server equipped with 384GB RAM, 10-core Intel(R) Xeon(R) Silver 4210R CPU @ 2.40GHz, and Nvidia RTX 3090 GPUs. 
The embedding dimensions for POIs and users are set to 10, and the time embedding dimension is set to 6. 
For the Transformer architecture, we incorporate two multi-head attention mechanisms and 2 encoder blocks. 
For the spatial decay rate $\beta$, we follow the settings of Flashback~\cite{yang2020location}.

\noindent\textbf{Overall Performance.} 
Table~\ref{tab:result} shows the predictive performance of all baseline methods and LoTNext on two datasets. 
Based on Table~\ref{tab:result}, we can draw the following conclusions:
\begin{itemize}[leftmargin=*]
\item On both public datasets, LoTNext outperforms all other state-of-the-art baseline methods across all metrics. 
Compared to the most recent and best-performing baseline method, SNPM, LoTNext achieves more significant improvements in Acc@1.
These results indicate that LoTNext is better at predicting long-tailed POIs that are less popular but highly relevant to specific users.

\item Utilizing graphs to model all user mobility patterns, thereby improving POI embeddings, representing as SNPM, Graph-Flashback, and GETNext, significantly outperform sequential methods represented by LSTPM and DeepMove, which rely solely on an individual user's short-term and long-term interests to predict the user's next location. 
However, the raw User-POI Interaction Graph has a large number of long-tailed nodes with a degree of 1 or very small. 
LoTNext, through its long-tailed graph adjustment module, effectively filters these long-tailed nodes, thereby enhancing the model's predictive performance.
\end{itemize}

\begin{table*}[!tpb]
\centering
\caption{Acc@k and MRR performance comparison on Gowalla and Foursquare datasets.}
\resizebox{\textwidth}{!}{
\begin{tabular}{l|cccc|cccc}
\toprule
\multirow{2}{*}{Model} & \multicolumn{4}{c|}{Gowalla}                                          & \multicolumn{4}{c}{Foursquare}                                        \\ \cline{2-9} 
                       & Acc@1           & Acc@5           & Acc@10          & MRR             & Acc@1           & Acc@5           & Acc@10          & MRR             \\ \midrule
ST-RNN~\cite{liu2016predicting}                 & 0.0900          & 0.2120          & 0.2730          & 0.1508          & 0.2290          & 0.4310          & 0.5050          & 0.3248          \\
DeepMove~\cite{feng2018deepmove}               & 0.0625          & 0.1304          & 0.1594          & 0.0982          & 0.2400          & 0.4319          & 0.4742          & 0.3270          \\
LBSN2Vec~\cite{yang2019revisiting}               & 0.0864          & 0.1186          & 0.1390          & 0.1032          & 0.2190          & 0.3955          & 0.4621          & 0.2781          \\
LightGCN~\cite{he2020lightgcn}               & 0.0428          & 0.1439          & 0.2115          & 0.1224          & 0.0540          & 0.1790          & 0.2710          & 0.1574          \\
LSTPM~\cite{sun2020go}                  & 0.0721          & 0.1843          & 0.2327          & 0.1306          & 0.2484          & 0.4489          & 0.5018          & 0.3365          \\
Flashback~\cite{yang2020location}              & 0.1158          & 0.2754          & 0.3479          & 0.1925          & 0.2496          & 0.5399          & 0.6326          & 0.3805          \\
STAN~\cite{luo2021stan}                   & 0.0891          & 0.2096          & 0.2763          & 0.1523          & 0.2265          & 0.4515          & 0.5310          & 0.3420          \\
GETNext~\cite{yang2022getnext}                & 0.1419          & 0.3270          & 0.4081          & 0.2294          & 0.2646          & 0.5640          & 0.6431          & 0.3988          \\
Graph-Flashback~\cite{rao2022graph}        & 0.1495          & 0.3399          & 0.4242          & 0.2401          & 0.2786          & 0.5733          & 0.6501          & 0.4109          \\
SNPM~\cite{yin2023next}                   & \underline{0.1593}    & \underline{0.3514}    & \underline{0.4346}    & \underline{0.2505}    & \underline{0.2899}    & \underline{0.5967}    & \underline{0.6763}    & \underline{0.4278}    \\ \midrule
\textbf{LoTNext (Ours) }                   & \textbf{0.1668} & \textbf{0.3605} & \textbf{0.4429} & \textbf{0.2591} & \textbf{0.3155} & \textbf{0.6059} & \textbf{0.6812} & \textbf{0.4469} \\ \bottomrule
\end{tabular}
}
\label{tab:result}
\end{table*}

\begin{figure}[!ht]
	\centering	
	\subfigure[Acc@1.]{
		\includegraphics[width=0.22\textwidth]{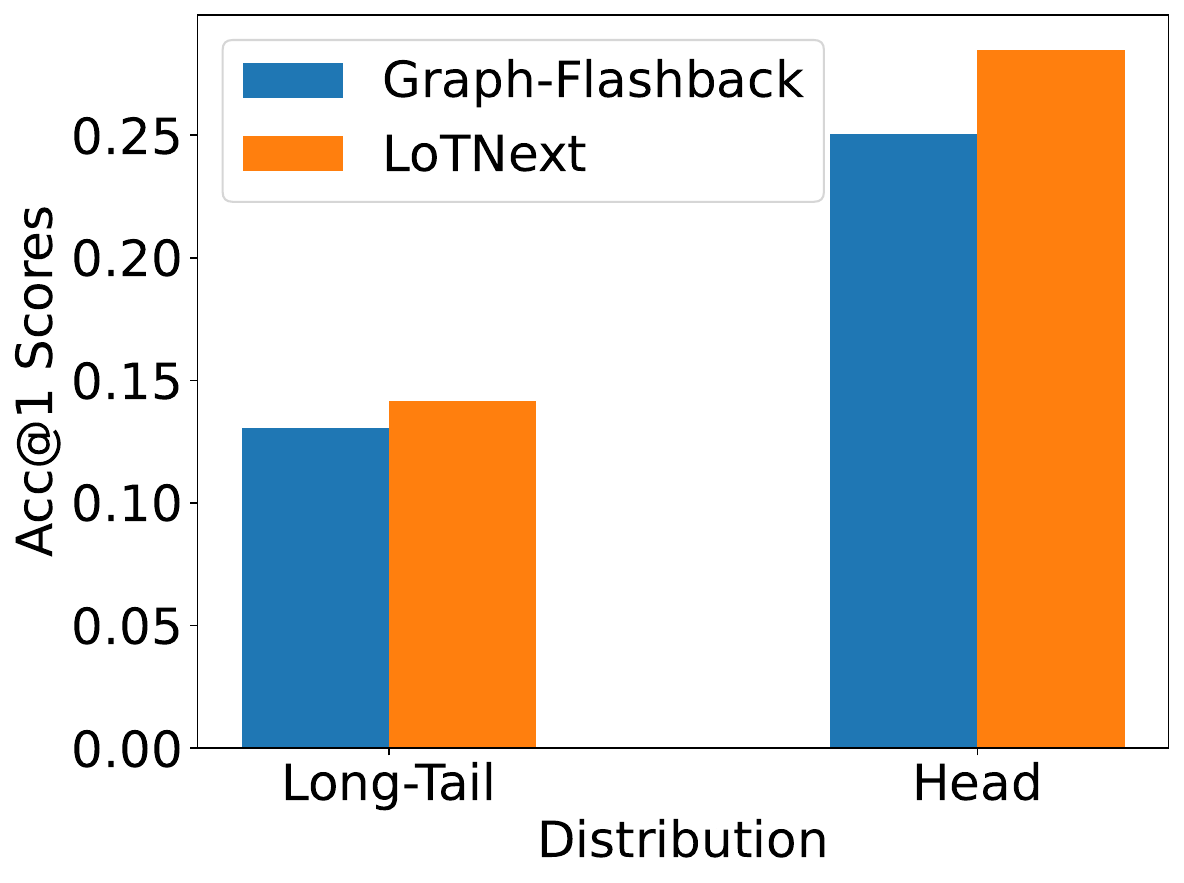}
		\label{fig:contrast_1}
	}\ \ \
	\subfigure[MRR.]{
		\includegraphics[width=0.22\textwidth]{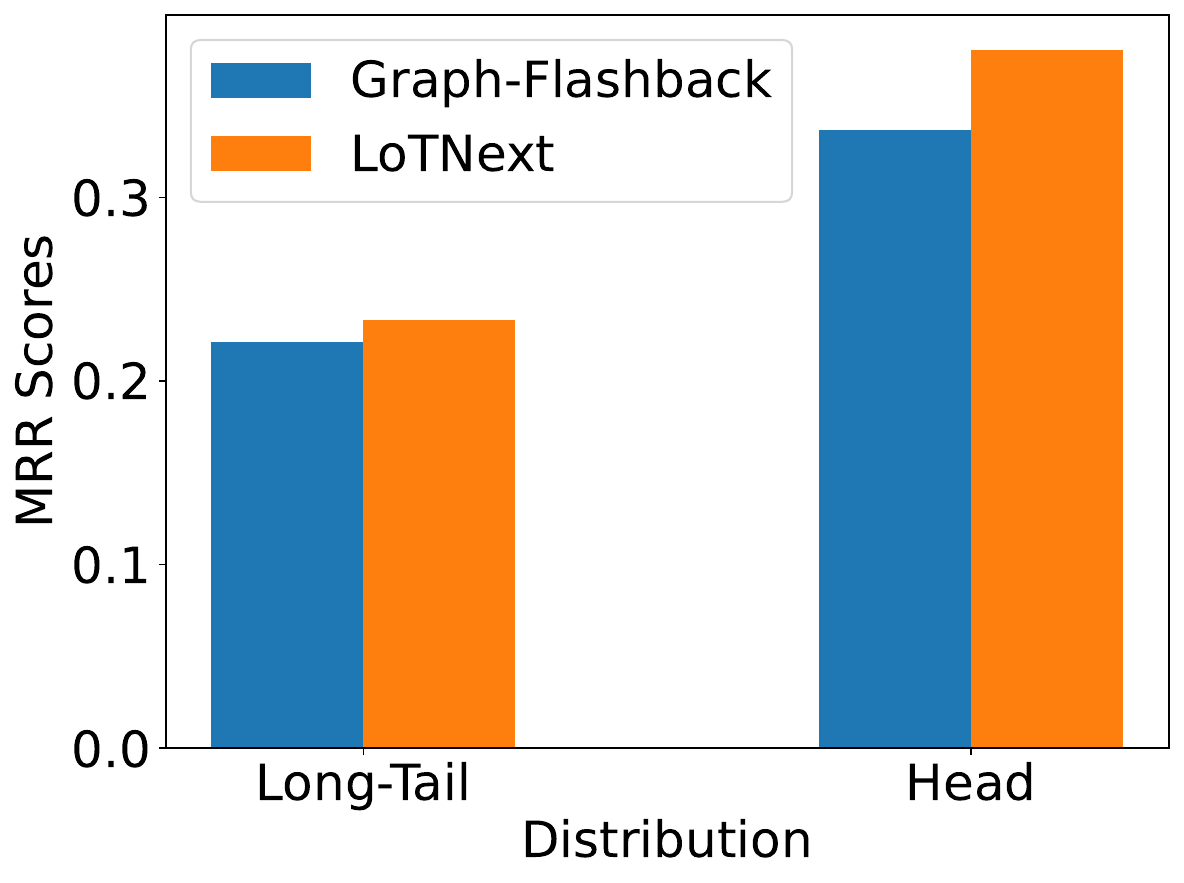}
		\label{fig:contrast_2} 
        }
  	\subfigure[Proportion of the predicted POIs.]{
        \includegraphics[width=0.42\textwidth]{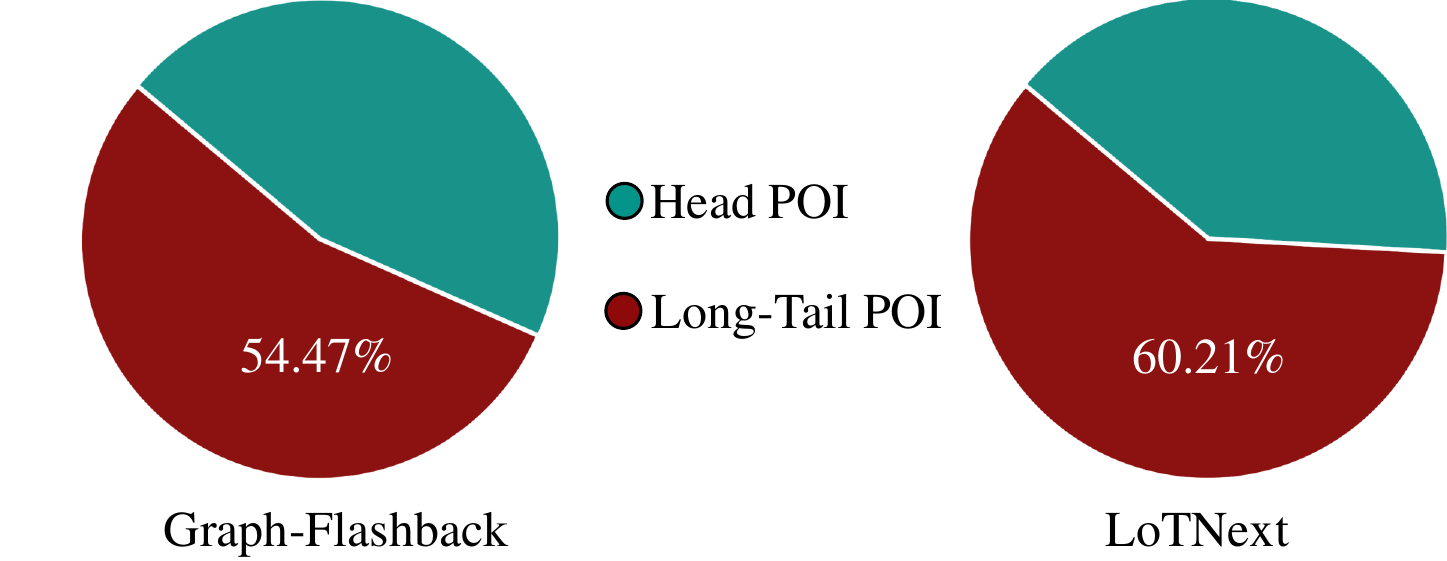}
		\label{fig:contrast_3} 
        } \
	\caption{The performance comparison of the long-tailed and head POIs between LoTNext and Graph-Flashback on Gowalla dataset.}
\label{fig:contrast}
\end{figure}

\noindent\textbf{Performance on Long-Tailed Samples.} To evaluate whether our model achieves accuracy improvement on long-tailed samples, we define samples with a frequency less than 100 on Gowalla dataset 
as long-tailed samples to test the model's specific predictive performance on both long-tailed and head samples. 
We compare LoTNext with Graph-Flashback which provides the pre-trained model for ease of comparison. 
As shown in Figure~\ref{fig:contrast_1} and Figure~\ref{fig:contrast_2}, LoTNext consistently outperforms Graph-Flashback on both Acc@1 and MRR metrics, whether for head or long-tailed samples. 
Furthermore, Figure~\ref{fig:contrast_3} reveals a notable distinction in the prediction of long-tailed POIs, LoTNext exhibits a roughly 6\% higher propensity to predict long-tailed POIs compared to Graph-Flashback. This increment not only underscores the enhanced capacity of LoTNext to identify and anticipate long-tailed POIs but also demonstrates the efficacy of our methodology. 

\noindent\textbf{Ablation Study.} 
To analyze the impact of different modules on LoTNext, we conducted the following ablation settings: (1) without the Long-Tailed Graph Adjustment module (w/o LTGA), where we conducted with the raw graph without graph adjustment. 
(2) without the Long-Tailed Loss Adjustment module (w/o LTLA), meaning we only used the original cross-entropy loss for testing. 
(3) without the original cross-entropy loss (w/o $\mathcal{L}_{CE}$), meaning we only use Long-Tailed Loss Adjustment module ($\mathcal{L}_{LTA}$ loss). 
(4) without the auxiliary prediction task module, we utilized the LTLA module and cross-entropy loss function, removing the auxiliary time prediction task, denoted as w/o $\mathcal{L}_{Aux}$.
\begin{table}[ht]
\small
\centering
\caption{The performance comparison among the LoTNext and variants without some components.}
\begin{tabular}{l|cccc|cccc}
\toprule
\multirow{2}{*}{Model} & \multicolumn{4}{c|}{Gowalla}    & \multicolumn{4}{c}{Foursquare}                                        \\ \cline{2-9} 
                       & Acc@1           & Acc@5           & Acc@10          & MRR        
 & Acc@1           & Acc@5           & Acc@10          & MRR  \\ \midrule
w/o LTGA                 & 0.1617          & 0.3568          & 0.4419          & 0.2550      & 0.3020          & 0.6002          & 0.6783          & 0.4368      \\
w/o LTLA               & 0.1544          & 0.3439          & 0.4266          & 0.2450   & 0.3014          & 0.5985          & 0.6758         & 0.4362         \\
w/o $\mathcal{L}_{CE}$                & 0.1550          & 0.3455          & 0.4287          & 0.2462   & 0.3029          & 0.5989          & 0.6771         & 0.4365         \\
w/o $\mathcal{L}_{Aux}$          & 0.1609          & 0.3567          & 0.4344          & 0.2523    & 0.3039          & 0.5993          & 0.6769          & 0.4370       \\ \midrule
\textbf{LoTNext}                  & \textbf{0.1668} & \textbf{0.3605} & \textbf{0.4429} & \textbf{0.2591} & \textbf{0.3155} & \textbf{0.6059} & \textbf{0.6812} & \textbf{0.4469} \\ \bottomrule
\end{tabular}

\label{tab:ablation}
\end{table}

From Table~\ref{tab:ablation}, we have the following findings: (1) The embeddings obtained after the LTGA module contribute to the model's predictive performance. 
This is mainly because long-tailed POIs can be considered noise to some extent, and appropriately eliminating some noise helps with model prediction. 
(2) Utilizing only the original cross-entropy loss results in performance below SNPM, indicating that the strategy of considering the long-tailed distribution through the LTLA module is effective for improving model accuracy in identifying the most relevant items. 
(3) The results using only the $\mathcal{L}_{LTA}$ loss show slightly higher metrics than results w/o LTLA, which suggests that the model may over-focus on long-tail data, leading to a decline in the recommendation performance for head data. For this reason, we consider incorporating the $\mathcal{L}_{CE}$ to balance the recommendation performance between long-tail data and head data.
(4) Without the time prediction task, we observe a decline in the MRR metric, suggesting that temporal features play a crucial role in helping the model capture the dynamic changes in user behavior.

\begin{figure}[!ht]
	\centering	
	\subfigure[Graph-Flashback.]{
		\includegraphics[width=0.4\textwidth]{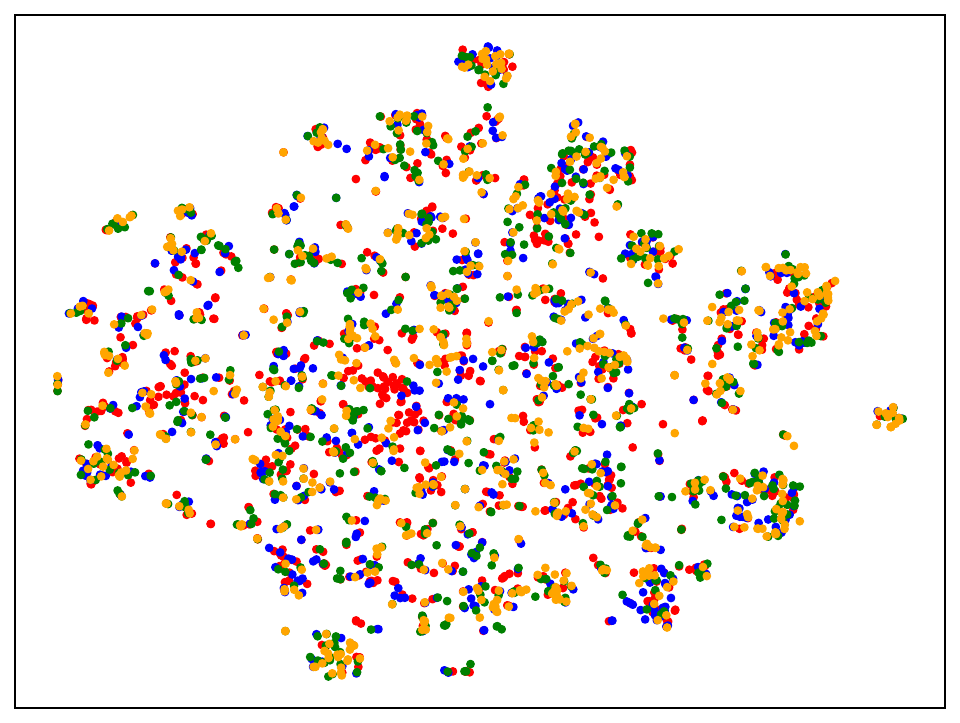}
		\label{fig:tsne_2}
	}\ \ \
 	\subfigure[LoTNext.]{
		\includegraphics[width=0.4\textwidth]{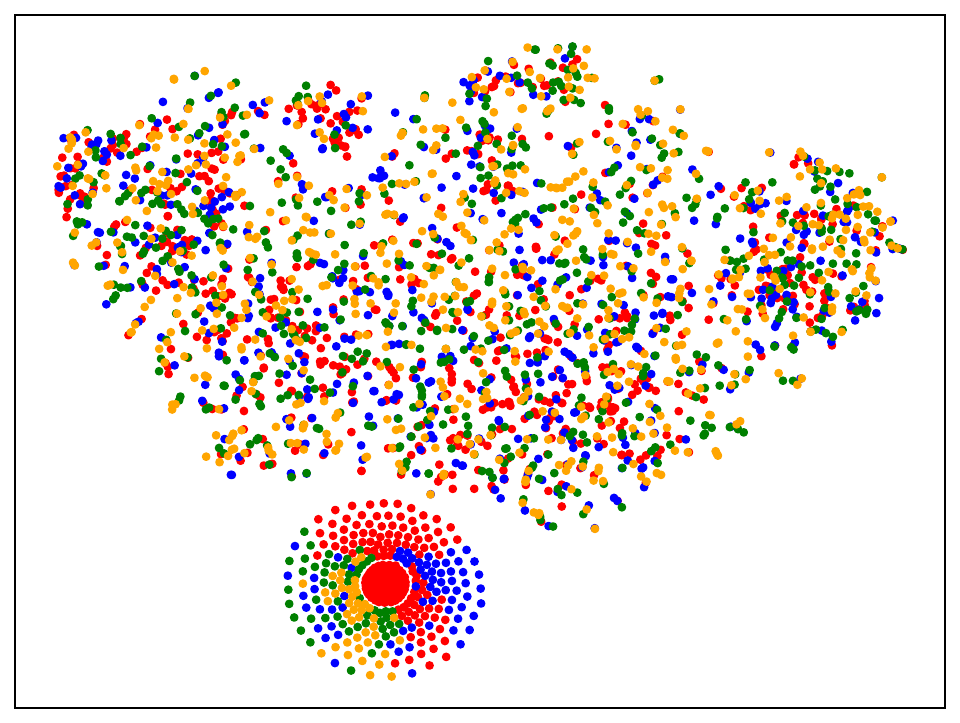}
		\label{fig:tsne_1}
	}\ \ \
	\caption{The visualization of tail POIs on Gowalla dataset. The color represents the POI frequency.}
 \label{fig:tsne}
\end{figure}

\noindent\textbf{Case Study: Learned POI Embedding. } 
Figure~\ref{fig:tsne} presents t-SNE visualizations of the embeddings for the four least frequently occurring POIs on Gowalla dataset. 
In Figure~\ref{fig:tsne_1} representing LoTNext the embeddings of these low-frequency POIs are more distinct and well-separated, indicating that LoTNext effectively captures the unique characteristics of these tail POIs.
This clear separation demonstrates that LoTNext can learn meaningful representations even for the least frequent POIs, which is crucial for accurate prediction and recommendation. 
In contrast, Figure~\ref{fig:tsne_2} showing Graph-Flashback's performance, reveals more overlapping and less distinct clusters for these low-frequency POIs. 
This overlap suggests that Graph-Flashback struggles to differentiate between the tail POIs, potentially leading to less accurate predictions for these rarely visited locations. 

\begin{wrapfigure}{r}{0.5\textwidth}
    \centering
    
    \includegraphics[width=0.5\textwidth]{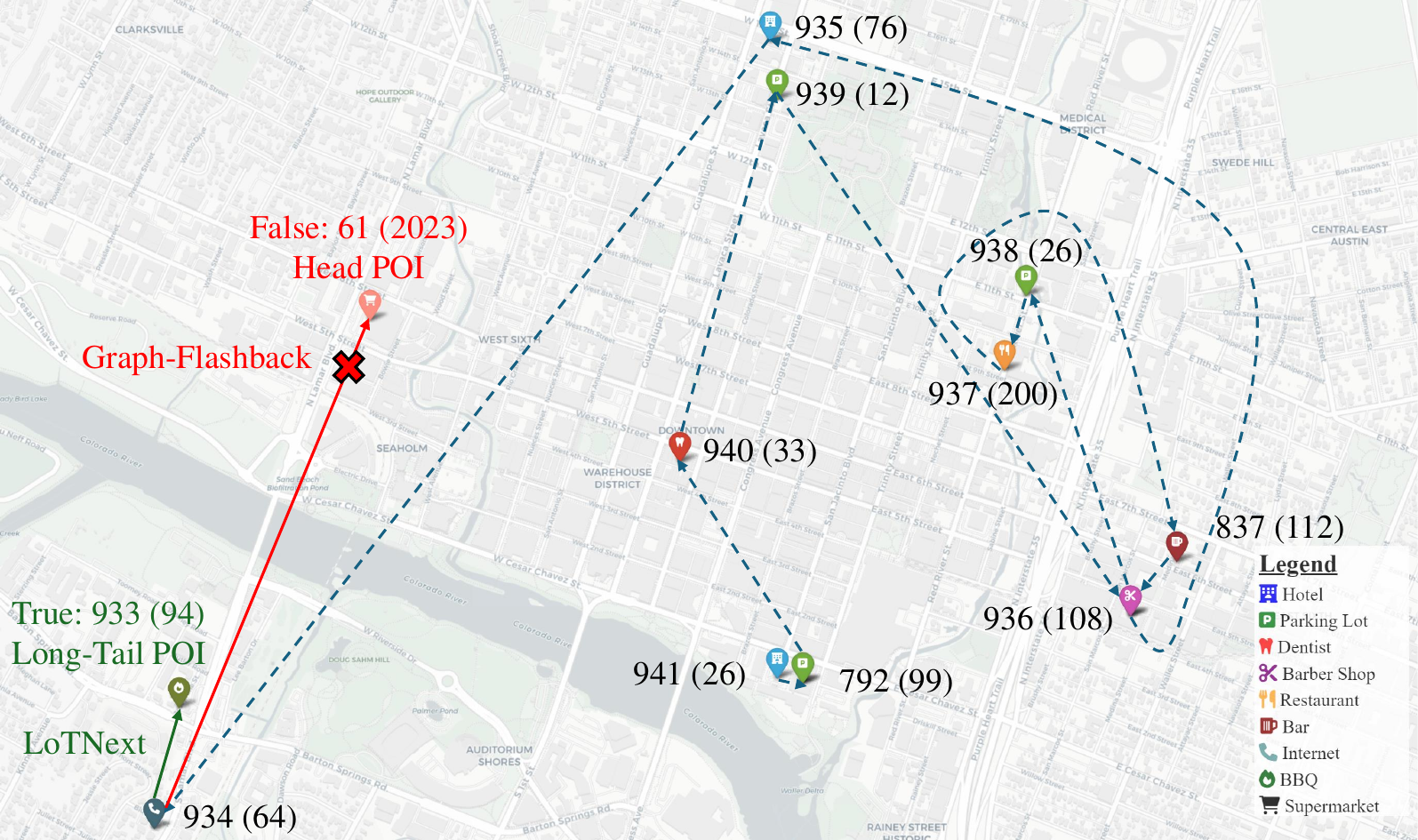}
    \vspace{-0.3cm}
    \caption{Sample prediction from Gowalla dataset with Graph-Flashback and LoTNext.}
    \label{fig:case}
    \vspace{-0.3cm}
\end{wrapfigure}
\noindent\textbf{\textls[-20]{Case Study: Prediction on Long-Tailed Sample.}} Figure~\ref{fig:case} provides a visual comparison of sample predictions made by the Graph-Flashback and LoTNext models on a trajectory from the Gowalla dataset for user 5. 
Each POI in the user's trajectory is identified by a unique ID and its visitation frequency, where the number in parentheses represents the frequency of visits.
In this specific trajectory, user 5 visits a sequence of POIs. For the given POI 934, LoTNext accurately predicts the next POI to be 933, a long-tail POI with a visitation frequency of 94. 
In contrast, the Graph-Flashback model incorrectly predicts the next POI to be 61, a head POI with an extremely high visitation frequency of 2023. 
This is the same sample as the problem shown in Figure~\ref{fig:longtail}, demonstrating the efficacy of LoTNext in capturing the user's actual movement pattern, which encompasses both frequently and infrequently visited POIs.

\section{Conclusion}
In this work, we propose LoTNext, a novel framework for human next POI prediction under long-tailed data distribution. 
Specifically, we employ a Long-Tailed Graph Adjustment module to mitigate the impact of long-tailed nodes within the User-POI Interaction Graph.
Additionally, to balance the influence of long-tailed data in the loss, we propose the Long-Tailed Loss Adjustment module to adjust the model's predicted logits and adaptively increase the weight of long-tailed samples. 
Moreover, we leverage the auxiliary prediction task to achieve spatial and temporal prediction synergy. 
Through comparisons with 10 state-of-the-art methods, we demonstrate the superiority of LoTNext over the most advanced approaches.
A limitation of our approach lies in that LoTNext's reliance on extensive user trajectory data poses a potential risk for privacy breaches if deployed by certain institutions or companies, which could lead to negative social impacts.
We plan to address it in future work.

\begin{ack}
This work was supported by JST SPRING Grant Number JPMJSP2108, JSPS KAKENHI Grant Number JP24K02996, JST CREST Grant Number JPMJCR21M2 including AIP challenge program, and Initiative on Recommendation Program for Young Researchers and Woman Researchers, Information Technology Center, The University of Tokyo.
\end{ack}

{
\small
\bibliographystyle{plainnat}
\bibliography{neurips_2024}

\begin{thebibliography}{53}
\providecommand{\natexlab}[1]{#1}
\providecommand{\url}[1]{\texttt{#1}}
\expandafter\ifx\csname urlstyle\endcsname\relax
  \providecommand{\doi}[1]{doi: #1}\else
  \providecommand{\doi}{doi: \begingroup \urlstyle{rm}\Url}\fi

\bibitem[Beutel et~al.(2017)Beutel, Chi, Cheng, Pham, and Anderson]{beutel2017beyond}
Alex Beutel, Ed~H Chi, Zhiyuan Cheng, Hubert Pham, and John Anderson.
\newblock Beyond globally optimal: Focused learning for improved recommendations.
\newblock In \emph{Proceedings of the 26th International Conference on World Wide Web}, pages 203--212, 2017.

\bibitem[Chen et~al.(2020)Chen, Jiang, Yang, Cai, Fan, Tsubouchi, Shibasaki, and Song]{chen2020dualsin}
Quanjun Chen, Renhe Jiang, Chuang Yang, Zekun Cai, Zipei Fan, Kota Tsubouchi, Ryosuke Shibasaki, and Xuan Song.
\newblock Dualsin: Dual sequential interaction network for human intentional mobility prediction.
\newblock In \emph{Proceedings of the 28th International Conference on Advances in Geographic Information Systems}, pages 283--292, 2020.

\bibitem[Du and Wu(2023)]{du2023no}
Yingxiao Du and Jianxin Wu.
\newblock No one left behind: Improving the worst categories in long-tailed learning.
\newblock In \emph{Proceedings of the IEEE/CVF Conference on Computer Vision and Pattern Recognition}, pages 15804--15813, 2023.

\bibitem[Fan et~al.(2018)Fan, Song, Xia, Jiang, Shibasaki, and Sakuramachi]{fan2018online}
Zipei Fan, Xuan Song, Tianqi Xia, Renhe Jiang, Ryosuke Shibasaki, and Ritsu Sakuramachi.
\newblock Online deep ensemble learning for predicting citywide human mobility.
\newblock \emph{Proceedings of the ACM on Interactive, Mobile, Wearable and Ubiquitous Technologies}, 2\penalty0 (3):\penalty0 1--21, 2018.

\bibitem[Fan et~al.(2022)Fan, Yang, Yuan, Jiang, Chen, Song, and Shibasaki]{fan2022online}
Zipei Fan, Xiaojie Yang, Wei Yuan, Renhe Jiang, Quanjun Chen, Xuan Song, and Ryosuke Shibasaki.
\newblock Online trajectory prediction for metropolitan scale mobility digital twin.
\newblock In \emph{Proceedings of the 30th International Conference on Advances in Geographic Information Systems}, pages 1--12, 2022.

\bibitem[Feng et~al.(2018)Feng, Li, Zhang, Sun, Meng, Guo, and Jin]{feng2018deepmove}
Jie Feng, Yong Li, Chao Zhang, Funing Sun, Fanchao Meng, Ang Guo, and Depeng Jin.
\newblock Deepmove: Predicting human mobility with attentional recurrent networks.
\newblock In \emph{Proceedings of the 2018 world wide web conference}, pages 1459--1468, 2018.

\bibitem[Gao et~al.(2022)Gao, Wang, Zhang, Yang, Miao, and Li]{gao2022self}
Qiang Gao, Wei Wang, Kunpeng Zhang, Xin Yang, Congcong Miao, and Tianrui Li.
\newblock Self-supervised representation learning for trip recommendation.
\newblock \emph{Knowledge-Based Systems}, 247:\penalty0 108791, 2022.

\bibitem[Han et~al.(2021)Han, Shang, Sun, Zhao, Zheng, and Zhang]{han2021point}
Peng Han, Shuo Shang, Aixin Sun, Peilin Zhao, Kai Zheng, and Xiangliang Zhang.
\newblock Point-of-interest recommendation with global and local context.
\newblock \emph{IEEE Transactions on Knowledge and Data Engineering}, 34\penalty0 (11):\penalty0 5484--5495, 2021.

\bibitem[He et~al.(2020)He, Deng, Wang, Li, Zhang, and Wang]{he2020lightgcn}
Xiangnan He, Kuan Deng, Xiang Wang, Yan Li, Yongdong Zhang, and Meng Wang.
\newblock Lightgcn: Simplifying and powering graph convolution network for recommendation.
\newblock In \emph{Proceedings of the 43rd International ACM SIGIR conference on research and development in Information Retrieval}, pages 639--648, 2020.

\bibitem[Huang et~al.(2019)Huang, Song, Fan, Jiang, Shibasaki, Zhang, Wang, and Kato]{huang2019variational}
Dou Huang, Xuan Song, Zipei Fan, Renhe Jiang, Ryosuke Shibasaki, Yu~Zhang, Haizhong Wang, and Yugo Kato.
\newblock A variational autoencoder based generative model of urban human mobility.
\newblock In \emph{2019 IEEE conference on multimedia information processing and retrieval (MIPR)}, pages 425--430. IEEE, 2019.

\bibitem[Jiang et~al.(2018{\natexlab{a}})Jiang, Song, Fan, Xia, Chen, Chen, and Shibasaki]{jiang2018deep}
Renhe Jiang, Xuan Song, Zipei Fan, Tianqi Xia, Quanjun Chen, Qi~Chen, and Ryosuke Shibasaki.
\newblock Deep roi-based modeling for urban human mobility prediction.
\newblock \emph{Proceedings of the ACM on Interactive, Mobile, Wearable and Ubiquitous Technologies}, 2\penalty0 (1):\penalty0 1--29, 2018{\natexlab{a}}.

\bibitem[Jiang et~al.(2018{\natexlab{b}})Jiang, Song, Fan, Xia, Chen, Miyazawa, and Shibasaki]{jiang2018deepurbanmomentum}
Renhe Jiang, Xuan Song, Zipei Fan, Tianqi Xia, Quanjun Chen, Satoshi Miyazawa, and Ryosuke Shibasaki.
\newblock Deepurbanmomentum: An online deep-learning system for short-term urban mobility prediction.
\newblock In \emph{Proceedings of the AAAI conference on artificial intelligence}, volume~32, 2018{\natexlab{b}}.

\bibitem[Jiang et~al.(2021)Jiang, Song, Fan, Xia, Wang, Chen, Cai, and Shibasaki]{jiang2021transfer}
Renhe Jiang, Xuan Song, Zipei Fan, Tianqi Xia, Zhaonan Wang, Quanjun Chen, Zekun Cai, and Ryosuke Shibasaki.
\newblock Transfer urban human mobility via poi embedding over multiple cities.
\newblock \emph{ACM Transactions on Data Science}, 2\penalty0 (1):\penalty0 1--26, 2021.

\bibitem[Jiang et~al.(2022)Jiang, Chen, Cai, Fan, Song, Tsubouchi, and Shibasaki]{jiang2022will}
Renhe Jiang, Quanjun Chen, Zekun Cai, Zipei Fan, Xuan Song, Kota Tsubouchi, and Ryosuke Shibasaki.
\newblock Will you go where you search? a deep learning framework for estimating user search-and-go behavior.
\newblock \emph{Neurocomputing}, 472:\penalty0 338--348, 2022.

\bibitem[Kim et~al.(2019)Kim, Kim, Park, and Yu]{kim2019sequential}
Yejin Kim, Kwangseob Kim, Chanyoung Park, and Hwanjo Yu.
\newblock Sequential and diverse recommendation with long tail.
\newblock In \emph{IJCAI}, volume~19, pages 2740--2746, 2019.

\bibitem[Kipf and Welling(2017)]{DBLP:conf/iclr/KipfW17}
Thomas~N. Kipf and Max Welling.
\newblock Semi-supervised classification with graph convolutional networks.
\newblock In \emph{5th International Conference on Learning Representations, {ICLR}}, 2017.

\bibitem[Lin et~al.(2017)Lin, Goyal, Girshick, He, and Doll{\'a}r]{lin2017focal}
Tsung-Yi Lin, Priya Goyal, Ross Girshick, Kaiming He, and Piotr Doll{\'a}r.
\newblock Focal loss for dense object detection.
\newblock In \emph{Proceedings of the IEEE international conference on computer vision}, pages 2980--2988, 2017.

\bibitem[Liu et~al.(2016)Liu, Wu, Wang, and Tan]{liu2016predicting}
Qiang Liu, Shu Wu, Liang Wang, and Tieniu Tan.
\newblock Predicting the next location: A recurrent model with spatial and temporal contexts.
\newblock In \emph{Proceedings of the AAAI conference on artificial intelligence}, volume~30, 2016.

\bibitem[Liu and Zheng(2020)]{liu2020long}
Siyi Liu and Yujia Zheng.
\newblock Long-tail session-based recommendation.
\newblock In \emph{Proceedings of the 14th ACM Conference on Recommender Systems}, pages 509--514, 2020.

\bibitem[Liu et~al.(2022)Liu, Yang, Xu, Yang, Huang, and Wang]{liu2022real}
Xin Liu, Yongjian Yang, Yuanbo Xu, Funing Yang, Qiuyang Huang, and Hong Wang.
\newblock Real-time poi recommendation via modeling long-and short-term user preferences.
\newblock \emph{Neurocomputing}, 467:\penalty0 454--464, 2022.

\bibitem[Luo et~al.(2023{\natexlab{a}})Luo, Ma, Xiao, and Song]{luo2023improving}
Sichun Luo, Chen Ma, Yuanzhang Xiao, and Linqi Song.
\newblock Improving long-tail item recommendation with graph augmentation.
\newblock In \emph{Proceedings of the 32nd ACM International Conference on Information and Knowledge Management}, pages 1707--1716, 2023{\natexlab{a}}.

\bibitem[Luo et~al.(2023{\natexlab{b}})Luo, Duan, Liu, and Chung]{luo2023timestamps}
Yan Luo, Haoyi Duan, Ye~Liu, and Fu-Lai Chung.
\newblock Timestamps as prompts for geography-aware location recommendation.
\newblock In \emph{Proceedings of the 32nd ACM International Conference on Information and Knowledge Management}, pages 1697--1706, 2023{\natexlab{b}}.

\bibitem[Luo et~al.(2021)Luo, Liu, and Liu]{luo2021stan}
Yingtao Luo, Qiang Liu, and Zhaocheng Liu.
\newblock Stan: Spatio-temporal attention network for next location recommendation.
\newblock In \emph{Proceedings of the web conference 2021}, pages 2177--2185, 2021.

\bibitem[Menon et~al.(2021)Menon, Jayasumana, Rawat, Jain, Veit, and Kumar]{MenonJRJVK21}
Aditya~Krishna Menon, Sadeep Jayasumana, Ankit~Singh Rawat, Himanshu Jain, Andreas Veit, and Sanjiv Kumar.
\newblock Long-tail learning via logit adjustment.
\newblock In \emph{9th International Conference on Learning Representations}, 2021.

\bibitem[Provost(2000)]{provost2000machine}
Foster Provost.
\newblock Machine learning from imbalanced data sets 101.
\newblock In \emph{Proceedings of the AAAI Workshop Imbalanced Data Sets}, pages 1--3, 2000.

\bibitem[Qin et~al.(2023)Qin, Wu, Ju, Luo, and Zhang]{qin2023diffusion}
Yifang Qin, Hongjun Wu, Wei Ju, Xiao Luo, and Ming Zhang.
\newblock A diffusion model for poi recommendation.
\newblock \emph{ACM Transactions on Information Systems}, 42\penalty0 (2):\penalty0 1--27, 2023.

\bibitem[Rao et~al.(2022)Rao, Chen, Liu, Shang, Yao, and Han]{rao2022graph}
Xuan Rao, Lisi Chen, Yong Liu, Shuo Shang, Bin Yao, and Peng Han.
\newblock Graph-flashback network for next location recommendation.
\newblock In \emph{Proceedings of the 28th ACM SIGKDD Conference on Knowledge Discovery and Data Mining}, pages 1463--1471, 2022.

\bibitem[Ren et~al.(2020)Ren, Yu, Ma, Zhao, Yi, et~al.]{ren2020balanced}
Jiawei Ren, Cunjun Yu, Xiao Ma, Haiyu Zhao, Shuai Yi, et~al.
\newblock Balanced meta-softmax for long-tailed visual recognition.
\newblock \emph{Advances in neural information processing systems}, 33:\penalty0 4175--4186, 2020.

\bibitem[Sun et~al.(2020)Sun, Qian, Chen, Liang, Nguyen, and Yin]{sun2020go}
Ke~Sun, Tieyun Qian, Tong Chen, Yile Liang, Quoc Viet~Hung Nguyen, and Hongzhi Yin.
\newblock Where to go next: Modeling long-and short-term user preferences for point-of-interest recommendation.
\newblock In \emph{Proceedings of the AAAI Conference on Artificial Intelligence}, volume~34, pages 214--221, 2020.

\bibitem[Tao et~al.(2023)Tao, Sun, Yang, Chen, Wang, Yang, Du, and Zheng]{tao2023local}
Yingfan Tao, Jingna Sun, Hao Yang, Li~Chen, Xu~Wang, Wenming Yang, Daniel Du, and Min Zheng.
\newblock Local and global logit adjustments for long-tailed learning.
\newblock In \emph{Proceedings of the IEEE/CVF International Conference on Computer Vision}, pages 11783--11792, 2023.

\bibitem[Tian et~al.(2020)Tian, Liu, Glaser, Hsu, and Kira]{tian2020posterior}
Junjiao Tian, Yen-Cheng Liu, Nathaniel Glaser, Yen-Chang Hsu, and Zsolt Kira.
\newblock Posterior re-calibration for imbalanced datasets.
\newblock \emph{Advances in Neural Information Processing Systems}, 33:\penalty0 8101--8113, 2020.

\bibitem[Vaswani et~al.(2017)Vaswani, Shazeer, Parmar, Uszkoreit, Jones, Gomez, Kaiser, and Polosukhin]{vaswani2017attention}
Ashish Vaswani, Noam Shazeer, Niki Parmar, Jakob Uszkoreit, Llion Jones, Aidan~N Gomez, {\L}ukasz Kaiser, and Illia Polosukhin.
\newblock Attention is all you need.
\newblock \emph{Advances in neural information processing systems}, 30, 2017.

\bibitem[Wang et~al.(2024)Wang, Jiang, Yang, Wu, Onizuka, Shibasaki, and Xiao]{wang2024large}
Jiawei Wang, Renhe Jiang, Chuang Yang, Zengqing Wu, Makoto Onizuka, Ryosuke Shibasaki, and Chuan Xiao.
\newblock Large language models as urban residents: An llm agent framework for personal mobility generation.
\newblock \emph{arXiv preprint arXiv:2402.14744}, 2024.

\bibitem[Wang et~al.(2023{\natexlab{a}})Wang, Fukumoto, Cui, Suzuki, Li, and Yu]{wang2023eedn}
Xinfeng Wang, Fumiyo Fukumoto, Jin Cui, Yoshimi Suzuki, Jiyi Li, and Dongjin Yu.
\newblock Eedn: Enhanced encoder-decoder network with local and global context learning for poi recommendation.
\newblock In \emph{Proceedings of the 46th International ACM SIGIR Conference on Research and Development in Information Retrieval}, pages 383--392, 2023{\natexlab{a}}.

\bibitem[Wang et~al.(2023{\natexlab{b}})Wang, Zhu, Wang, Ma, Li, and Yu]{wang2023adaptive}
Zhaobo Wang, Yanmin Zhu, Chunyang Wang, Wenze Ma, Bo~Li, and Jiadi Yu.
\newblock Adaptive graph representation learning for next poi recommendation.
\newblock In \emph{Proceedings of the 46th International ACM SIGIR Conference on Research and Development in Information Retrieval}, pages 393--402, 2023{\natexlab{b}}.

\bibitem[Wei et~al.(2023)Wei, Liang, Liu, Dai, Li, and Wang]{wei2023meta}
Chunyu Wei, Jian Liang, Di~Liu, Zehui Dai, Mang Li, and Fei Wang.
\newblock Meta graph learning for long-tail recommendation.
\newblock In \emph{Proceedings of the 29th ACM SIGKDD Conference on Knowledge Discovery and Data Mining}, pages 2512--2522, 2023.

\bibitem[Wu et~al.(2020{\natexlab{a}})Wu, Li, Zhao, and Qian]{wu2020personalized}
Yuxia Wu, Ke~Li, Guoshuai Zhao, and Xueming Qian.
\newblock Personalized long-and short-term preference learning for next poi recommendation.
\newblock \emph{IEEE Transactions on Knowledge and Data Engineering}, 34\penalty0 (4):\penalty0 1944--1957, 2020{\natexlab{a}}.

\bibitem[Wu et~al.(2020{\natexlab{b}})Wu, Pan, Chen, Long, Zhang, and Philip]{wu2020comprehensive}
Zonghan Wu, Shirui Pan, Fengwen Chen, Guodong Long, Chengqi Zhang, and S~Yu Philip.
\newblock A comprehensive survey on graph neural networks.
\newblock \emph{IEEE transactions on neural networks and learning systems}, 32\penalty0 (1):\penalty0 4--24, 2020{\natexlab{b}}.

\bibitem[Xu et~al.(2023)Xu, Suzumura, Yong, Hanai, Yang, Kanezashi, Jiang, and Fukushima]{xu2023revisiting}
Xiaohang Xu, Toyotaro Suzumura, Jiawei Yong, Masatoshi Hanai, Chuang Yang, Hiroki Kanezashi, Renhe Jiang, and Shintaro Fukushima.
\newblock Revisiting mobility modeling with graph: A graph transformer model for next point-of-interest recommendation.
\newblock In \emph{Proceedings of the 31st ACM International Conference on Advances in Geographic Information Systems}, pages 1--10, 2023.

\bibitem[Xue et~al.(2021)Xue, Salim, Ren, and Oliver]{xue2021mobtcast}
Hao Xue, Flora Salim, Yongli Ren, and Nuria Oliver.
\newblock Mobtcast: Leveraging auxiliary trajectory forecasting for human mobility prediction.
\newblock \emph{Advances in Neural Information Processing Systems}, 34:\penalty0 30380--30391, 2021.

\bibitem[Yan et~al.(2023)Yan, Song, Jiao, He, Wang, Li, and Chu]{yan2023spatio}
Xiaodong Yan, Tengwei Song, Yifeng Jiao, Jianshan He, Jiaotuan Wang, Ruopeng Li, and Wei Chu.
\newblock Spatio-temporal hypergraph learning for next poi recommendation.
\newblock In \emph{Proceedings of the 46th international ACM SIGIR conference on research and development in information retrieval}, pages 403--412, 2023.

\bibitem[Yang et~al.(2019)Yang, Qu, Yang, and Cudre-Mauroux]{yang2019revisiting}
Dingqi Yang, Bingqing Qu, Jie Yang, and Philippe Cudre-Mauroux.
\newblock Revisiting user mobility and social relationships in lbsns: a hypergraph embedding approach.
\newblock In \emph{The world wide web conference}, pages 2147--2157, 2019.

\bibitem[Yang et~al.(2020)Yang, Fankhauser, Rosso, and Cudre-Mauroux]{yang2020location}
Dingqi Yang, Benjamin Fankhauser, Paolo Rosso, and Philippe Cudre-Mauroux.
\newblock Location prediction over sparse user mobility traces using rnns.
\newblock In \emph{Proceedings of the Twenty-Ninth International Joint Conference on Artificial Intelligence}, pages 2184--2190, 2020.

\bibitem[Yang et~al.(2022{\natexlab{a}})Yang, Jiang, Song, and Guo]{yang2022survey}
Lu~Yang, He~Jiang, Qing Song, and Jun Guo.
\newblock A survey on long-tailed visual recognition.
\newblock \emph{International Journal of Computer Vision}, 130\penalty0 (7):\penalty0 1837--1872, 2022{\natexlab{a}}.

\bibitem[Yang et~al.(2022{\natexlab{b}})Yang, Liu, and Zhao]{yang2022getnext}
Song Yang, Jiamou Liu, and Kaiqi Zhao.
\newblock Getnext: trajectory flow map enhanced transformer for next poi recommendation.
\newblock In \emph{Proceedings of the 45th International ACM SIGIR Conference on research and development in information retrieval}, pages 1144--1153, 2022{\natexlab{b}}.

\bibitem[Yin et~al.(2023)Yin, Liu, Shen, Chen, Shang, and Han]{yin2023next}
Feiyu Yin, Yong Liu, Zhiqi Shen, Lisi Chen, Shuo Shang, and Peng Han.
\newblock Next poi recommendation with dynamic graph and explicit dependency.
\newblock In \emph{Proceedings of the AAAI Conference on Artificial Intelligence}, volume~37, pages 4827--4834, 2023.

\bibitem[Yin et~al.(2012)Yin, Cui, Li, Yao, and Chen]{YinCLYC12}
Hongzhi Yin, Bin Cui, Jing Li, Junjie Yao, and Chen Chen.
\newblock Challenging the long tail recommendation.
\newblock \emph{Proc. {VLDB} Endow.}, 5\penalty0 (9):\penalty0 896--907, 2012.

\bibitem[Yu et~al.(2020)Yu, Cui, Guo, Lu, Li, and Lu]{yu2020category}
Fuqiang Yu, Lizhen Cui, Wei Guo, Xudong Lu, Qingzhong Li, and Hua Lu.
\newblock A category-aware deep model for successive poi recommendation on sparse check-in data.
\newblock In \emph{Proceedings of the web conference 2020}, pages 1264--1274, 2020.

\bibitem[Zhang et~al.(2020)Zhang, Sun, Zhang, Kloeden, and Klanner]{zhang2020modeling}
Lu~Zhang, Zhu Sun, Jie Zhang, Horst Kloeden, and Felix Klanner.
\newblock Modeling hierarchical category transition for next poi recommendation with uncertain check-ins.
\newblock \emph{Information Sciences}, 515:\penalty0 169--190, 2020.

\bibitem[Zhang et~al.(2023)Zhang, Kang, Hooi, Yan, and Feng]{zhang2023deep}
Yifan Zhang, Bingyi Kang, Bryan Hooi, Shuicheng Yan, and Jiashi Feng.
\newblock Deep long-tailed learning: A survey.
\newblock \emph{IEEE Transactions on Pattern Analysis and Machine Intelligence}, 2023.

\bibitem[Zhang et~al.(2021)Zhang, Cheng, Yao, Yi, Hong, and Chi]{zhang2021model}
Yin Zhang, Derek~Zhiyuan Cheng, Tiansheng Yao, Xinyang Yi, Lichan Hong, and Ed~H Chi.
\newblock A model of two tales: Dual transfer learning framework for improved long-tail item recommendation.
\newblock In \emph{Proceedings of the web conference 2021}, pages 2220--2231, 2021.

\bibitem[Zhao et~al.(2022)Zhao, Chen, Tan, Huang, and Zhu]{zhao2022adaptive}
Yan Zhao, Weicong Chen, Xu~Tan, Kai Huang, and Jihong Zhu.
\newblock Adaptive logit adjustment loss for long-tailed visual recognition.
\newblock In \emph{Proceedings of the AAAI Conference on Artificial Intelligence}, volume~36, pages 3472--3480, 2022.

\bibitem[Zhu et~al.(2023)Zhu, Ye, Zhang, Zhao, and Yu]{zhu2023difftraj}
Yuanshao Zhu, Yongchao Ye, Shiyao Zhang, Xiangyu Zhao, and James Yu.
\newblock Difftraj: Generating gps trajectory with diffusion probabilistic model.
\newblock \emph{Advances in Neural Information Processing Systems}, 36:\penalty0 65168--65188, 2023.

\end{thebibliography}
}

\appendix

\section{Appendix / supplemental material}

\subsection{Notations}
The notations used in our paper are summarized as follows.
\begin{table}[!ht]
\small
\centering
\caption{Notation Table.}
\label{tab:notation}
\begin{tabular}{l|l}
\toprule
Symbol                                                                           & Meaning                                                              \\ \midrule
$U, u$                                                                             & User set and user                                                    \\
$P, p$                                                                             & POI set and POI                                                      \\
$T, t$                                                                                & Time slot set and time slot                                         \\
$E^U, E^P, E^T, E_{pos}$                                                                 & Embedding of user, POI, timestamp, and position \\
$X, \widetilde{X}$                                                                            & Embedding sequence w/o./with positional embedding                 \\
$Z, \widetilde{Z}$ & Output of the multi-head attention layer and transformer \\ 
$\widetilde{Z}', \widetilde{z}'$ & Refined output by spatial contextual attention layer \\ 
$\mathcal{O}, o$ & Input of the final fully connected layer \\ 
$L$ & Logit scores calculated by fully connected layer \\
$W, w$                                                                                & Trainable weight matrix                                              \\
$N$                                                                                & The number of sample                                                 \\
$B$                                                                                & Batchsize                                         \\
$\mathcal{G}^{In}$                                                            & User-POI Interaction Graph                                                 \\
$\mathcal{G}^{Tr}$                                                            & Global Transition Graph                                                   \\
$\mathcal{V}^{In}, \mathcal{A}^{In}$                                                            & Input node feature matrix and adjacent matrix of $\mathcal{G}^{In}$                                                \\
$\mathcal{V}^{Tr}, \mathcal{A}^{Tr}$                                                            & Input node feature matrix and adjacent matrix of $\mathcal{G}^{Tr}$                                                    \\
$A_{ij}$                                                                                & Attention scores                                                 \\
$\widetilde{\mathcal{G}}^{In}, \widetilde{\mathcal{A}}^{In}$                                                                                & Refined User-POI Interaction Graph and its adjacent matrix                                              \\
$E^{In}, E^{Tr}$                                                                 & Node embedding of $\mathcal{G}^{In}$ and $\mathcal{G}^{Tr}$\\
$D^{In}, D^{Tr}$                                                                 & Degree matrix of $\mathcal{G}^{In}$ and $\mathcal{G}^{Tr}$\\
$\widetilde{E}^P$                                                                 & Denoised POI embedding \\
\midrule
$d_{p}$, $d_{t}, d_u$                                                                         & Hidden dimension of the POI, time and user                                      \\
$n$                                                                                & Sequence length                                                     \\
$b$                                                                                & Bias                                                                 \\
$y_{i}$                                                                        & Ground-truth
indicator                         \\
$l, l'$                                                                            & Logits and adjusted logits for each POI class                                          \\
$t_i, \hat{t}_i$                                                & Time slot and the forecasted time slot                          \\ 
\midrule
$\omega$                                                              & Spatial weight                                                   \\
$\beta$                                                             & Distance decay weight                                            \\
$\Delta(i, j)$                                                & Haversine distance between $p_i$ and $p_j$                        \\
$\epsilon$                                                          & Small constant               \\
$\sigma$                                                            & Sigmoid function                                                     \\
$\delta$                                                            & Denoising threshold                                                   \\
$\tau$                                                              & Logit adjustment weight                                                   \\
$\alpha$                                                              & Adjustment factor                                                  \\ 
$\xi, \widetilde{\xi}$                                                              & Adjusted vector magnitude and its geometric mean                                     \\
$\phi$                                                           & Adaptive weights for each sample                                           \\
$\lambda$                                                           & Learnable loss weights                                           \\
 \bottomrule
\end{tabular}
\end{table}

\subsection{Computational Cost}
In this section, we explore the computational cost of LoTNext. We selected three sequence-based and three graph-based baselines to demonstrate the computational efficiency of our approach. Table~\ref{tab:compute} lists the inference time for each deep learning model during the testing phase (running one training/testing instance, i.e., test time divided by batch size). We ensured that all models were executed on the same RTX 3090 GPU. Surprisingly, due to batch training, the graph-based methods generally run significantly faster than the sequence-based methods. DeepMove is the fastest among the sequence-based methods, as it only considers calculating attention using historical trajectories. Compared to DeepMove, LSTPM further introduces a geographical relationship adjacency matrix to enrich the spatial context, making it slightly slower than DeepMove. STAN employs a dual-layer attention architecture, with one attention layer aggregating spatiotemporal correlations within user trajectories and the other selecting the most likely next POI based on weighted check-ins, resulting in the longest inference time for STAN.

\begin{table}[!ht]
\caption{Comparison of computational cost. Each method is benchmarked on the same NVIDIA
GeForce RTX 3090 GPU.}
\label{tab:compute}
\centering
\begin{tabular}{cc}
\toprule
\multicolumn{1}{c|}{Method}          & Inference Time ($10^{-3}$ Seconds) \\ \midrule
\multicolumn{1}{c|}{DeepMove (Sequence-based)}        & 1.422                         \\
\multicolumn{1}{c|}{LSTPM (Sequence-based)}           & 3.417                         \\
\multicolumn{1}{c|}{STAN (Sequence-based)}            & 2887.809                       \\
\multicolumn{1}{c|}{GETNext (Graph-based)}         & 3.824                         \\
\multicolumn{1}{c|}{Graph-Flashback (Graph-based)} & 0.0918                        \\
\multicolumn{1}{c|}{SNPM (Graph-based)}            & 0.491                              \\ \midrule
\multicolumn{1}{c|}{LoTNext (Graph-based)}                              & 0.257                         \\ \bottomrule
\end{tabular}
\end{table}

In the graph-based methods, GETNext introduces additional computational overhead due to the need for extra POI candidate probability reorganization based on transition attention during the final prediction stage. SNPM requires extra computation time due to the search for similar neighborhoods within the graph. As for our LoTNext, it requires more time to run compared to Graph-Flashback because LoTNext includes graph denoising and an auxiliary temporal prediction task. However, Table~\ref{tab:result} and Table~\ref{tab:ablation} demonstrate the effectiveness of our proposed modules, even at the cost of some computational time. Thus, considering that LoTNext encompasses more processing steps and overall accuracy, the increase in inference time is still acceptable.

\subsection{Hyperparameter Analysis}
\begin{figure}[!ht]
	\centering	
	\subfigure[The impact of parameter $\delta$ on Gowalla dataset.]{
		\includegraphics[width=0.22\textwidth]{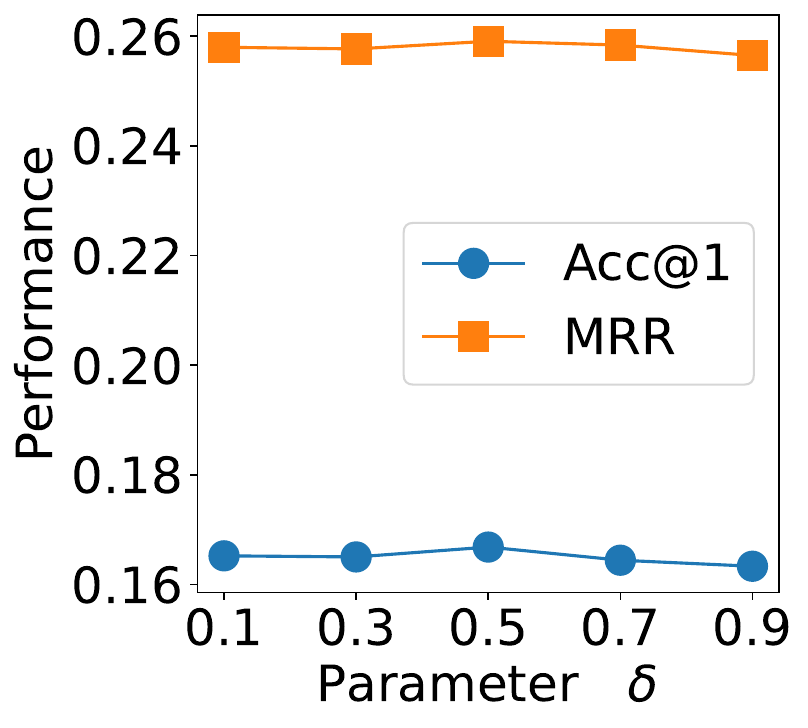}
		\label{fig:hyper_1}
	}\ \ \
	\subfigure[The impact of parameter $\tau$ on Gowalla dataset.]{
		\includegraphics[width=0.22\textwidth]{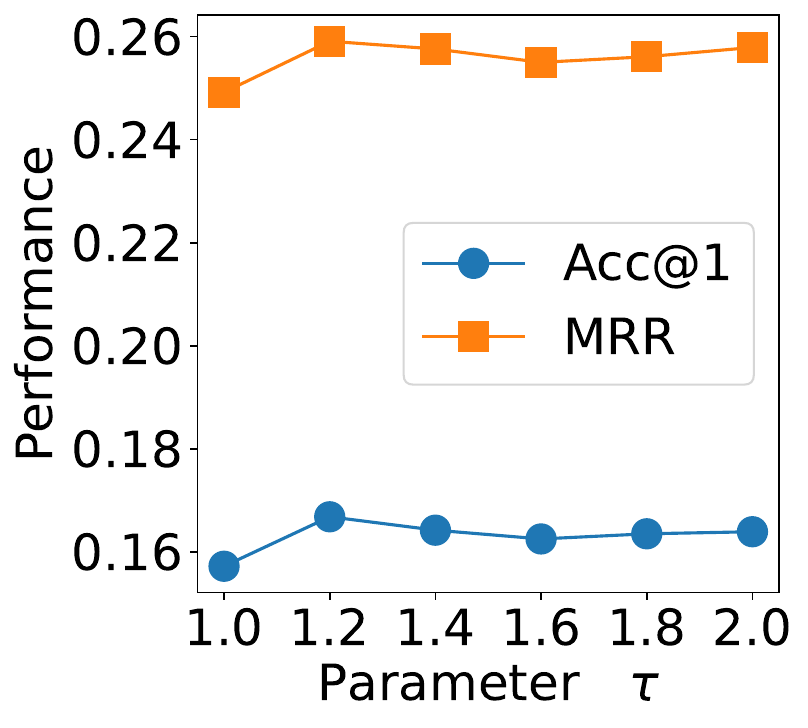}
		\label{fig:hyper_2}
	}
        \subfigure[The impact of parameter $\delta$ on Foursquare dataset.]{
		\includegraphics[width=0.22\textwidth]{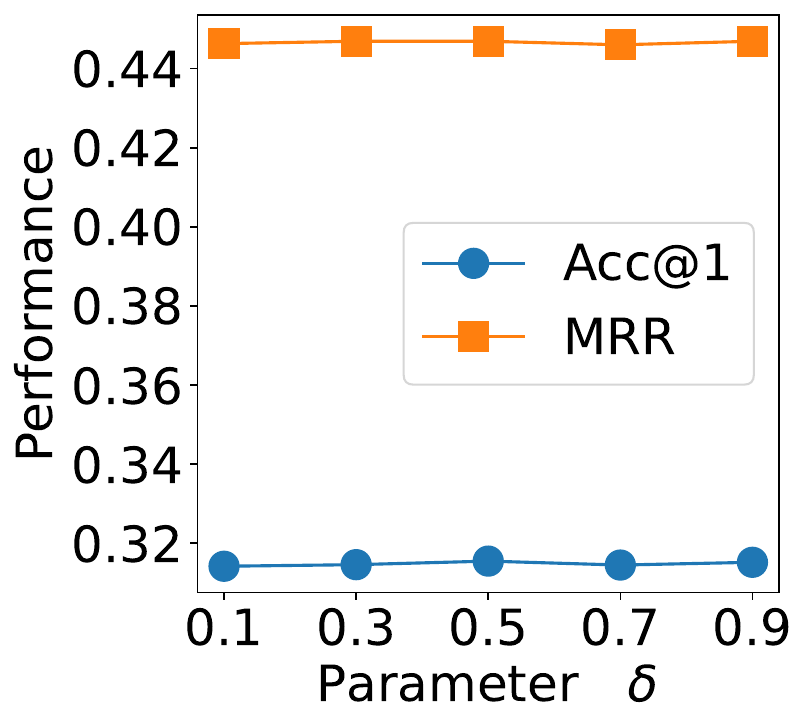}
		\label{fig:hyper_3}
	}\ \ \
	\subfigure[The impact of parameter $\tau$ on Foursquare dataset.]{
		\includegraphics[width=0.22\textwidth]{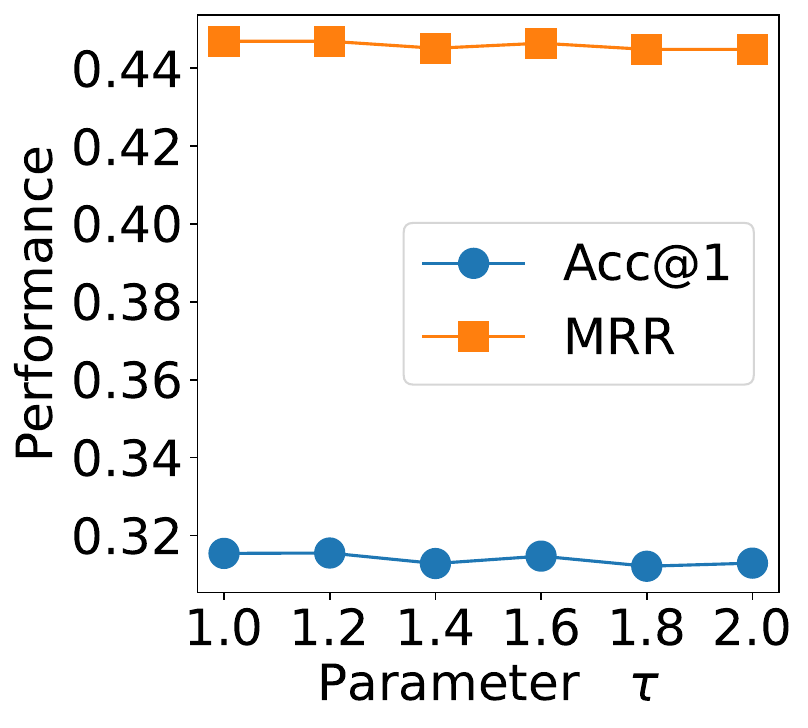}
		\label{fig:hyper_4}
	}
	\caption{Impact of denoising thresholds $\delta$ and logit adjustment weight $\tau$.}
 \label{fig:hyper}
\end{figure}
We conduct hyperparameter sensitivity experiments on the Long-Tailed Graph Adjustment module's threshold \(\delta\) and the weight \(\tau\) of the logit adjustment module to identify the optimal parameter values on  Gowalla and Foursquare datasets. 
We first experiment with a range of thresholds \(\delta\) from 0.1 to 0.9 in increments of 0.2, which controls the sensitivity of the model to the long-tailed distribution by filtering less significant edges in the graph. 
The results, shown in Figure~\ref{fig:hyper_1} for Gowalla and Figure~\ref{fig:hyper_3} for Foursquare, indicate that Acc@1 and MRR remain stable across different values, with the optimal threshold identified as \(\delta = 0.5\). 
Next, we vary the logit adjustment weight \(\tau\) from 1 to 2 in increments of 0.2 to test the model's performance in balancing class imbalances. 
Figure~\ref{fig:hyper_2} and Figure~\ref{fig:hyper_4} reveal that \(\tau = 1.2\) yields the best results on both datasets, suggesting a moderate adjustment weight helps generalize better without overly amplifying rare classes. 
These consistent findings across both datasets underscore the robustness of \(\delta = 0.5\) and \(\tau = 1.2\), highlighting the importance of hyperparameter tuning in improving model accuracy and ranking metrics for better prediction of user behavior in diverse datasets.

\subsection{Model Training Pseudo-code}
Algorithm~\ref{alg:LoTNext} shows the pseudo-code of the LoTNext training process. In our experiments, all training instances are processed through mini-batches.
\begin{algorithm}
\small
\caption{Pseudo-code of training LoTNext}
\label{alg:LoTNext}
\begin{algorithmic}[1]
\STATE \textbf{Input:} User set $U$, POI set $P$, user check-in sequences $Q_u$ for each user $u \in U$
\STATE \textbf{Output:} Trained model parameters $\gamma$
\STATE $\gamma$ $\leftarrow$ Initialize randomly

\WHILE{not converge}
    \STATE Construct graph $G^{In}$ and $G^{Tr}$, apply denoising, and learn node embeddings $E^{In}$ and $E^{Tr}$ by Eq.~(\ref{eq:1})-(\ref{eq:2_2}) \hfill $\triangleright$ Graph Adjustment
    \STATE Compute denoised POI embedding $\tilde{E}^P$ based on $E^{In}$ and $E^{Tr}$  \hfill $\triangleright$ Embedding Denoising
    
    \STATE Calculate time and user embeddings $E^T$, $E^U$, and construct embedding sequence $X$ \hfill $\triangleright$ Embedding Initialization
    \STATE Transform input $X$ to $\tilde{X}$ \hfill $\triangleright$ Positional Encoding
    \STATE Calculate Transformer encoder output $\tilde{Z}$ by Eq.~(\ref{eq:3}) \hfill $\triangleright$ Transformer Encoder

    \FOR{each POI $p_k$ in sequence}
        \STATE Calculate spatial weight $\omega_k$ by Eq.~(\ref{eq:5}) \hfill $\triangleright$ Spatial Weight Calculation
        \STATE Refine output $\tilde{z}_k$ by Eq.~(\ref{eq:6}) \hfill $\triangleright$ Output Refinement
    \ENDFOR

    \STATE Calculate fused output $O$ and logits $L$  \hfill $\triangleright$ Prediction Layer
    \STATE Compute cross-entropy loss $L_{CE}$ by Eq.~(\ref{eq:18_2}) \hfill $\triangleright$ Loss Calculation

    \STATE Adjust logits using $\alpha_i$ and recompute logits $\tilde{l}_i$ by Eq.~(\ref{eq:10}) \hfill $\triangleright$ Logit Adjustment
    \STATE Calculate adaptive weights $\phi^k$ and overall loss $L_{LTA}$ by Eq.~(\ref{eq:12})-(\ref{eq:15_2}) \hfill $\triangleright$ Loss Adjustment
    
    \STATE Compute auxiliary loss $L_{Aux}$ by Eq.~(\ref{eq:17}) \hfill $\triangleright$ Auxiliary Loss
    \STATE Update parameters $\gamma$ by minimizing joint loss $L_{Joint}$ by Eq.~(\ref{eq:19}) \hfill $\triangleright$ Parameter Update
\ENDWHILE

\STATE \textbf{return} $\gamma$
\end{algorithmic}
\end{algorithm}

\end{document}